\documentclass[11pt]{article}
\usepackage[a4paper, margin=1in]{geometry}
\usepackage{graphicx}
\usepackage{cite}
\usepackage{amsmath}
\usepackage{hyperref}
\usepackage{url}
\usepackage{amsfonts}
\usepackage{eurosym}
\usepackage{longtable}
\usepackage{xcolor}
\usepackage{float}
\usepackage{longtable}
\usepackage{tabularx}
\usepackage{enumitem}
\usepackage{lscape}
\usepackage{changepage}
\usepackage{tabularx}
\usepackage{longtable}
\usepackage{booktabs}
\usepackage{tabularx}  
\usepackage{xcolor}
\usepackage{rotating}

\usepackage{algorithm}
\usepackage{algorithmicx}
\usepackage{algcompatible}
\renewcommand{\COMMENT}[1]{\hfill \% #1}

\usepackage{quantikz}

\usepackage{amsmath}

\usepackage{libertinus}

\usepackage{cancel}
\usepackage{authblk}
\usepackage[protrusion=true,expansion=true]{microtype}
\SetProtrusion
   [ name     = mysettings,  
     factor   = 1000 ]      
   { encoding = *, family = * }
   { , = {500,500}, . = {250,250}, - = {300,300}, ? = {300,300}, ! = {300,300} }

\usepackage{hyperref}
\hypersetup{
    colorlinks=true,
    linkcolor=blue,
    citecolor=blue,
    filecolor=magenta,
    urlcolor=cyan
}

\newif\ifdraft
\drafttrue

\ifdraft
\newcommand{\fsnote}[1]{ {\textcolor{orange} { ***Francesc: #1 }}}
\newcommand{\crnote}[1]{ {\textcolor{blue} { ***Carlos: #1 }}}
\newcommand{\jknote}[1]{ {\textcolor{red} { ***Johannes: #1 }}}
\newcommand{\gjnote}[1]{ {\textcolor{purple} { ***Geert-Jan: #1 }}}
\newcommand{\ycnote}[1]{\textcolor{green!50!black}{***Yudong: #1}}
\newcommand{\metanote}[1]{ {\textcolor{orange} { #1 }}}
\newcommand{\marvnote}[1]{ {\textcolor{cyan} { #1 }}}

\else
\newcommand{\fsnote}[1]{}
\newcommand{\crnote}[1]{}
\newcommand{\jknote}[1]{}
\newcommand{\gjnote}[1]{}
\newcommand{\ycnote}[1]{}
\newcommand{\metanote}[1]{}
\newcommand{\marvnote}[1]{}
\fi

\title{Towards solving industrial integer linear programs with Decoded Quantum Interferometry 
}
\author[1,3]{Francesc Sabater}
\author[2]{Ouns El Harzli}
\author[2]{Geert-Jan Besjes}
\author[1]{Marvin Erdmann}
\author[1]{Johannes Klepsch}
\author[2]{Jonas Hiltrop}
\author[2]{Jean-François Bobier}
\author[2]{Yudong Cao \thanks{cao.yudong@bcg.com}}
\author[1]{Carlos A. Riofrío \thanks{carlos.riofrio@bmwgroup.com}}
\affil[1]{BMW Group, Munich, Germany}
\affil[2]{Boston Consulting Group and BCG X AI Science Institute, Boston, USA}
\affil[3]{Technical University of Munich, TUM School of Natural Sciences, Physics Department, 85748 Garching, Germany}

\date{\today}

\begin{document}
\maketitle

\begin{abstract}
Optimization via decoded quantum interferometry (DQI) has recently gained a great deal of attention as a promising avenue for solving optimization problems using quantum computers. In this paper, we apply DQI to an industrial optimization problem in the automotive industry: the vehicle option-package pricing problem. Our main contributions are 1) formulating the industrial problem as an integer linear program (ILP), 2) converting the ILP into instances of max-XORSAT, and 3) developing  a detailed quantum circuit implementation for belief propagation, a heuristic algorithm for decoding LDPC codes. Thus, we provide a full implementation of the DQI algorithm using Belief Propagation, which can be applied to any industrially relevant ILP by first transforming it into a max-XORSAT instance. We also evaluate the effectiveness of our implementation by benchmarking it against both Gurobi and a random sampling baseline.

\end{abstract}

\tableofcontents
\newpage

\section{Introduction}

\textbf{The search for quantum advantage in hard combinatorial optimization problems.}
Decoded Quantum Interferometry (DQI) \cite{jordan2024optimizationdecodedquantuminterferometry} emerges recently from a growing body of evidence that Fourier-space, intrinsically quantum “decoding” can unlock algorithmic structure that has resisted both classical heuristics and popular variational approaches \cite{ChaillouxTillich2024QDP,DebrisAlazardFallahpourStehle2024STOC,DebrisAlazardRemaudTillich2024TIT,ChenEtAl2024LWEAmplitudes,ChaillouxTillich2024SoftDecoders}. On the one hand, the filtering framework of Chen, Liu and Zhandry \cite{ChenLiuZhandry2021} shows that, given appropriately prepared quantum LWE-like states, discrete Fourier transforms and amplitude filtering yield polynomial-time algorithms for variants of average-case lattice problems (including LWE-state solving and Extended Dihedral Coset Problem), thereby demonstrating the concrete power of quantum interference to decode hidden linear structure. On the other hand, while the Quantum Approximate Optimization Algorithm (QAOA) remains a leading heuristic, a sequence of rigorous results has exposed structural obstacles—from symmetry/locality barriers and concentration phenomena to constant-depth limitations on high-girth graphs and spin models—tempering hopes for broad, worst-case separations in its native regime \cite{BravyiEtAl2020Obstacles,BarakMarwaha2022HighGirth,AnshuMetger2023QAOALimits,BassoEtAl2022FOCS}, even as there is promising empirical scaling evidence on specific hard families (e.g., LABS) \cite{ShaydulinEtAl2024SciAdv}. A separate rigorous line of inquiry shows provable quantum speedups for constraint satisfaction problems (CSP) via quantum-walk acceleration of backtracking trees, but just like applying Grover's algorithm for exhaustively searching over the solution space for NP-complete problems, these gains are essentially quadratic in a tree-size parameter and do not directly translate into approximation guarantees \cite{Montanaro2018Backtracking}. DQI advances a third route \cite{jordan2024optimizationdecodedquantuminterferometry}: it uses the quantum Fourier transform to convert objective evaluation into syndrome information and then reduces optimization to decoding, so that algorithmic progress hinges on the strength of the inserted decoder (classical today, quantum tomorrow) rather than on low-depth variational expressivity. In particular, DQI provides instance-wise, decoder-controlled guarantees and recasts problems such as max-XORSAT as LDPC-style decoding, thereby importing decades of coding-theoretic machinery while explicitly leveraging the same interference-for-decoding principle that underlies the filtering results established previously \cite{ChenLiuZhandry2021}. This decoding-centric paradigm thus offers a rigorous and complementary path in the search for quantum advantage in optimization—one that sidesteps known variational barriers while capitalizing on quantum Fourier structure where it is provably effective.

\textbf{Pervasive importance of optimization in industry.} A complementary component in the search for quantum advantage in practice is the structure of optimization problems arising from industrial settings. NP-hard optimization problems are pervasive in industry, and a very large fraction can be (and routinely are) modeled as 0–1 Integer Linear Programs (ILPs). Canonical examples include day-ahead unit commitment in power systems, where binary on/off decisions for generators are scheduled under network and ramping constraints \cite{Knueven2020UC}; airline crew pairing and related tail-assignment problems built as set-partitioning ILPs with column generation \cite{Anbil1998CrewPairing}; facility-location and $p$-median decisions with 0–1 site-opening variables \cite{Cornuejols1983UFLP}; cutting-stock and bin-packing formulations using 0–1 pattern-selection variables \cite{GilmoreGomory1961CuttingStock}; and vehicle-routing families whose strongest exact methods still rely on 0–1 formulations, branch-and-cut, and set-partitioning relaxations \cite{TothVigo2002VRP}. These applications frequently embed hard “kernel” structures: Set Cover, max-SAT, or Quadratic Assignment—identified as NP-complete/NP-hard in classic work \cite{Karp1972,SahniGonzalez1976QAP}, yet the real instances also exhibit exploitable sparsity, locality, decomposability, and business rules that worst-case reductions ignore. As a result, industrial-grade solvers blend branch-and-bound with powerful cutting planes and decades of engineering (e.g., SCIP and commercial MIP engines), often augmented with meta-heuristics \cite{Bixby2012HistoryMIP,Achterberg2009SCIP}; and in the satisfiability world, modern MaxSAT frameworks leverage SAT oracles via core-guided and hitting-set methods \cite{DaviesBacchus2011MaxHS,Morgado2013IterativeCoreGuided}. Against this backdrop, the significance of the present paper is to show a principled pipeline that maps general 0–1 ILPs to ILP-constraint feasibility, and then—via linear/pseudo-Boolean encodings long studied in the SAT community \cite{EenSorensson2006PBtoSAT,Sinz2005Cardinality,BailleuxBoufkhad2003Totalizer,RousselManquinho2009HB} —to weighted max-XORSAT instances whose parity-check structure is tailor-made for DQI’s Fourier-space decoding. This “ILP $\rightarrow$ feasibility $\rightarrow$ max-XORSAT” route both aligns with established SAT/PB-encoding practice and makes the LDPC-like algebraic structure explicit, enabling the DQI reductions \cite{jordan2024optimizationdecodedquantuminterferometry} to be applied to realistic business ILPs; the companion dynamic-pricing study provides a concrete end-to-end demonstration of this mapping on an automotive bundling/pricing ILP.

\textbf{Benefits of expanding on implementation details.} Finally, to move from promise to practice we must turn abstract algorithms into concrete, logical circuits: only gate-level designs expose the true cost profile (T-count/T-depth, ancilla footprint, space–time volume) and the hidden assumptions (state preparation, data-loading, and oracle realizations) that determine whether a putative speedup survives fault tolerance and hardware constraints \cite{Preskill2018NISQ,GidneyEkera2019RSA,GidneyFowler2019Factories,Babbush2018LinearT,vonBurg2021PRR}. Detailed circuitizations have repeatedly changed conclusions—e.g., factoring’s end-to-end resource was cut by two orders of magnitude once modular arithmetic, QROM, and factory scheduling were engineered explicitly \cite{GidneyEkera2019RSA}, while chemistry “linear-T” qubitization results tie optimistic asymptotics to million-qubit-and-hours regimes only after full block-encoding circuits are costed \cite{Babbush2018LinearT,vonBurg2021PRR}. This level of specificity also anchors fault-tolerant planning: non-Clifford throughput must be matched by magic-state factories with quantified layout/latency, making factory design a first-class part of the algorithm \cite{GidneyFowler2019Factories}; rotation-synthesis limits translate precision targets into T-gate budgets \cite{RossSelinger2014ZRot}. Crucially, pedagogical clarity and reproducibility follow when algorithms are delivered as machine-readable circuits/IR—OpenQASM 3 and QIR serve as shared substrates for compilation, scheduling, and verification across toolchains and architectures \cite{OpenQASM3,QIRSpec}. In short, concrete circuits are not a nicety but the prerequisite for credible quantum-advantage claims: they turn asymptotic sketches into testable engineering artifacts, support apples-to-apples comparisons, and provide the basis for full fault-tolerant resource estimates and co-design.

\textbf{Related works and our contribution.} After the initial DQI paper \cite{jordan2024optimizationdecodedquantuminterferometry}, subsequent work by Patamawisut et al \cite{Patamawisut2025DQICircuit} supplied the first end-to-end quantum circuitization of DQI’s decoder via a reversible Gauss–Jordan elimination (GJE) design, addressing the core challenge of operating a decoder coherently on superpositions and giving gate-depth/ancilla analyses on small instances—thereby providing a concrete, algebraic alternative to BP within the DQI paradigm. Most recently, Bu et al \cite{Bu2025DQINoise} analyzed DQI under local noise and identified a noise-weighted sparsity parameter governing performance degradation on OPI and Max-XORSAT, offering guidance for algorithm-and-instance co-design under realistic error models. This work extends DQI along two complementary axes: (i) it provides a systematic industrial pipeline from 0–1 ILPs to Max-XORSAT with explicit construction of parity-check matrices $B$ engineered for LDPC-like structure, thereby bridging abstract DQI instances to real business constraints; and (ii) it implements a coherent, binary hard-decision BP decoder (BP1) as a quantum circuit integrated into the full DQI flow (state preparation$\rightarrow$phase/syndrome 
$\rightarrow$encoding$\rightarrow$coherent decoding $\rightarrow$Hadamard$\rightarrow$measurement), alongside benchmarking on an automotive option-package pricing ILP against Gurobi and random baselines, logical resource estimates, small-scale simulations, and an appendix contrasting a GJE decoder within the same framework. 

The rest of paper is organized as follows: in Sec.~\ref{sec:dqi}, we briefly summarize the DQI algorithm. In Sec.~\ref{sec:implementation_bp1}, we detail our quantum circuit implementation of the decoding step of DQI. In Sec.~\ref{sec:problem_formulation}, we formulate the industrial optimization problem. In Sec.~\ref{sec: trasnformation to max-XOR-Sat} we discuss how to transform the presented  optimization problem or any other ILP to a max-XORSAT instance. In Sec.~\ref{sec:methods}, we describe the methods we follow for our numerical experiments validating our quantum circuit implementation, and in Sec.~\ref{sec: results} we detail our findings. Finally, we conclude the paper in Sec.~\ref{sec:conclusions} with a summary of our conclusions and a discussion of possible future directions.

\section{Decoded Quantum Interferometry (DQI)} \label{sec:dqi}

DQI \cite{jordan2024optimizationdecodedquantuminterferometry} is a newly proposed algorithm to solve optimization problems on a quantum computer by leveraging quantum interference patterns and classical decoding techniques. In a nutshell, DQI converts optimization problems into decoding problems and uses the quantum Fourier transform (QFT) to amplify the probability of measuring high-quality solutions. To this effect, the algorithm prepares quantum states that are more likely to yield solutions with high objective value when measured. In this section, we briefly describe the algorithmic steps to prepare the DQI state for binary ILPs which are mapped to max-XORSAT problems. 

A general 0-1 ILP is defined as follows:
\begin{equation}\label{eq:ilp_objective}
\min_{\mathbf{z}\in\{0,1\}^{n'}}\mathbf{c}^T\mathbf{z},
\end{equation}
subject to
\begin{equation}\label{eq:ilp_constraints}
     C\mathbf{z} \geq\mathbf{a} \\
\end{equation}
where $\mathbf{z} \in \{0,1\}^{n'}$ is the binary-valued decision variable of dimension $n'$, and $\mathbf{c} \in \mathbb{R}^{n'}$, $C \in \mathbb{R}^{m' \times n'}$, and $\mathbf{a} \in \mathbb{R}^{m'}$ are the parameters of the optimization problem. This problem is then transformed into a XOR satisfiability problem in which the objective is to maximize the number of fulfilled constraints (here all operations are mod 2)
\begin{equation}\label{eq:xorsat_constraints}
    B\mathbf{x}=\mathbf{v}
\end{equation}
where $\mathbf{x}\in\{0,1\}^n$ is the decision variable,  $\mathbf{v}\in\{0,1\}^m$, and $B^T\in\{0,1\}^{n\times m}$ encodes constraints of the satisfiability problem and defines the parity check matrix used later by DQI. We provide details of the transformation in Sec.~\ref{sec: trasnformation to max-XOR-Sat}. The problem is thus
\begin{equation}\label{eq:max_xorsat_objective}
    \max_{\mathbf{x}\in\ \{0,1\}^{n}} \sum_{i=1}^m(-1)^{v_i+\mathbf{b}_i\mathbf{x}}
\end{equation}
where $\mathbf{b}_i$ is the $i$th row of $B$, and, as before, all operations in the exponent are mod 2. Eq.~\eqref{eq:max_xorsat_objective} computes the maximum difference between satisfied and unsatisfied XOR constraints. In general, the transformation from the problem defined in Eqs.~\eqref{eq:ilp_objective} and \eqref{eq:ilp_constraints} to its max-XORSAT formulation Eqs.~\eqref{eq:xorsat_constraints} and \eqref{eq:max_xorsat_objective} incurs in an overhead which implies that $n>n'$ and $m>m'$. In the rest of the section, we describe the DQI algorithm for a general max-XORSAT problem defined by $B$ and $\mathbf{v}$. We closely follow \cite{jordan2024optimizationdecodedquantuminterferometry}, but conveniently refer to the quantum registers using classical error correction jargon.

\subsection{Dicke-state preparation}

    The algorithm starts by preparing the \emph{message} register, consisting of $m$ qubits, in the state
    \begin{equation}
        \label{eq: Dicke state}
        |\Psi_0\rangle = \sum_{k=0}^{\ell} w_k \frac{1}{\sqrt{\binom{m}{k}}} \sum_{|\mathbf{y}| = k} |\mathbf{y}\rangle,
    \end{equation}
    where $\sum_{|\mathbf{y}| = k} |\mathbf{y}\rangle$ denotes the uniform superposition over all computational basis states of Hamming weight $k$, and $\ell$ is the maximum number of errors the decoder is required to decode. This register encodes the bit strings,  representing code words corrupted by bit flips, that need to be decoded. This normalized superposition is known as a Dicke state of weight $k$ over $m$ qubits~\cite{Bartschi_2022,Wang_2024}
    \begin{equation}
        |D_{m,k}\rangle = \frac{1}{\sqrt{\binom{m}{k}}} \sum_{|\mathbf{y}| = k} |\mathbf{y}\rangle.
    \end{equation}
    Following the prescription in Ref.~\cite{jordan2024optimizationdecodedquantuminterferometry}, the coefficients $w_k \in \mathbb{R}^{\ell+1}$ are given as the components of the principal eigenvector of a $(\ell+1) \times (\ell+1)$ tridiagonal matrix $A$ given by:
    \begin{equation}
    A =
    \begin{pmatrix}
    0 & a_1 & 0 & \cdots & 0 \\
    a_1 & d & a_2 & \cdots & 0 \\
    0 & a_2 & 2d & \cdots & 0 \\
    \vdots & \vdots & \vdots & \ddots & a_\ell \\
    0 & 0 & 0 & a_\ell & \ell d
    \end{pmatrix},
    \end{equation}
    where the parameters are defined as
    \begin{equation}
        a_k = \sqrt{k(m - k + 1)}, \quad \quad d = \frac{p - 2r}{\sqrt{r(p - r)}}.
    \end{equation}
    For max-XORSAT problems, $p = 2$ and $r = 1$.

\subsection{Encoding of problem-specific phases}
The next step is to encode the appropriate phases into the \emph{message} register applying a $Z$ gate to each qubit, $i$, for which $v_i=1$. We get then the quantum state 
\begin{equation}
\label{eq: encoding phases}
    |\Psi_1\rangle = \prod_{i=1}^m Z_i^{v_i}|\Psi_0\rangle = \sum_{k=0}^{\ell} w_k \frac{1}{\sqrt{\binom{m}{k}}} 
    \sum_{|\mathbf{y}| = k} (-1)^{\mathbf{v} \cdot \mathbf{y}} 
    \left| \mathbf{y} \right\rangle.
\end{equation}

\subsection{Encoding of the syndrome}
Next, we create a \textit{syndrome} register of $n$ qubits where we encode $B^T\mathbf{y}$. The quantum state that we obtain at this stage is then 
\begin{equation}
    |\Psi_2\rangle=U|\Psi_1\rangle|0^{\otimes n}\rangle = \sum_{k=0}^{\ell} w_k \frac{1}{\sqrt{\binom{m}{k}}} 
    \sum_{|\mathbf{y}| = k} (-1)^{\mathbf{v} \cdot \mathbf{y}} 
    \left| \mathbf{y} \right\rangle \left| B^T \mathbf{y} \right\rangle
\end{equation}
where $U$ is an entangling unitary, which we will describe in Sec. \ref{sec:quantum_circuit}.

\subsection{Decoding}
The next step is to uncompute the \emph{message} register. That is, for each $\mathbf{y}$ in the superposition,
\begin{equation}
    \label{eq: decoding simple goal}
    U_D|\mathbf{y}\rangle \left| B^T \mathbf{y} \right\rangle \rightarrow |0^{\otimes m}\rangle \left| B^T \mathbf{y} \right\rangle,
\end{equation}
we aim to transform $\mathbf{y}$ back to the all-zero state, while preserving the $B^T \mathbf{y}$ component, where $U_D$ is the unitary that achieves the uncomputation. If the decoder succeeds, we can write the state of the system as 
\begin{equation}\label{eq:decoded_state}
    |\Psi_3\rangle = \sum_{k=0}^{\ell} w_k \frac{1}{\sqrt{\binom{m}{k}}} 
    \sum_{|\mathbf{y}| = k} (-1)^{\mathbf{v} \cdot \mathbf{y}} 
    \left| B^T \mathbf{y} \right\rangle.
\end{equation}
This is the most challenging and central step in the DQI algorithm. The authors of Ref.~\cite{jordan2024optimizationdecodedquantuminterferometry} identify this task to be the well-known problem of \emph{syndrome decoding} in classical error correction. Here, the matrix $B^T$ defines a binary linear code, e.g., a low density parity-check (LDPC) code, and transforming $\mathbf{y} \rightarrow 0^{\otimes m}$ based solely on knowledge of $B^T \mathbf{y}$ is equivalent to decoding the error from a given syndrome.

\subsection{Hadamard Transform}
In the final step, the \emph{message} register is discarded and a Hadamard gate is applied to each qubit of the \emph{syndrome} register. If the decoding procedure succeeds, this operation prepares the desired DQI state 
\begin{equation}\label{eq:dqi_state}
    |DQI\rangle = \prod_{i=1}^n H_i|\Psi_3\rangle.
\end{equation}
This final state has the property that the probability to measure the optimal solution of the optimization problem is enhanced.

\subsection{Measurement and Postprocessing}
In practice, we measure all qubits in both the \emph{syndrome} and \emph{message} registers in the computational basis. Since the decoder may not succeed with 100\% probability, we post-process the measurement outcomes by keeping only those instances where the \emph{message} register is measured in the all-zero state. This ensures that the \emph{message} register has been properly uncomputed. The measurement results from the \emph{syndrome} register encode candidate solutions to the max-XORSAT problem under consideration.

\section{Quantum implementation of a binary belief propagation decoder}\label{sec:implementation_bp1}
\subsection{Binary belief propagation (BP1)}\label{sec:gallager_bf}
In this section, we present a simple, multi-purpose binary belief propagation (BP1) algorithm, which we later implement coherently as a quantum circuit. This algorithm is based on the original work~\cite{1057683}, in which LDPCs were introduced. This algorithm is sometimes referred to as Gallager-type or Gallager A/B belief propagation \cite{10.5555/1795974, PhysRevLett.84.1355}. A revisited theory of good error correction codes and their decoders can be found in Ref.~\cite{748992}. Algorithm \ref{alg:gallager_bp} shows the pseudo code of the BP1 algorithm. This algorithm takes as input the parity check matrix $B^T$, a bit string $\mathbf{y}$ corresponding to some code word that was damaged during transmission through a noisy channel, and a maximum number of iterations $T$. In this formulation, the algorithm tries to decode a \emph{damaged} all-zero code word, $\mathbf{y}$, back to $\mathbf{c}=[0,\ldots,0]^T$. The algorithm succeeds only when it returns the all-zero word $\mathbf{c}$. This is a more stringent requirement than returning $\mathbf{b}$ such that $B^T\mathbf{b}\mod 2=0$. The more stringent case is necessary for the DQI implementation, as we need to coherently map code words containing bit flips, $|\mathbf{y}\rangle$, to the all-zero state $|0^{\otimes m}\rangle$ as described in the previous section.

This algorithm computes the error syndrome $B^T\mathbf{y}$, counts the number of ones in the syndrome which are connected to a given entry of $\mathbf{y}$ and decides to flip that entry based on a given threshold. In this implementation, we have chosen the threshold to be the number of connected nodes, which are defined as the cardinality of the set of indices $i$ such that $B^T_{ij} = 1$. In other words, the threshold is the number of participating code-word bits in the parity check. This choice of threshold empirically performed the best in our experiments, i.e., it maximized the success rate of the decoder. Notably, this algorithm chooses which entries to flip based on a hard threshold. Generally, this reduces the efficiency of the decoding, but in our case, it also simplifies the algorithm enough to make a quantum circuit implementation feasible.

\begin{figure}[ht]
    \centering
    \includegraphics[width=0.9\linewidth]{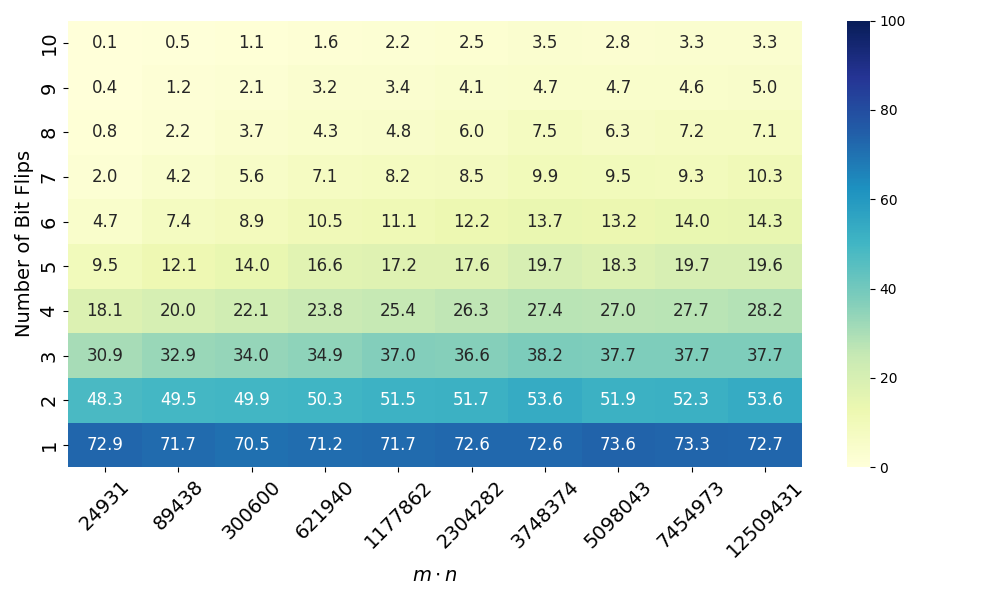}
    \caption{Success rate of our implementation of the BP1 decoder for the zero code-word for $\ell$ randomly generated  bit flips as a function of $\ell$ and the problem size, i.e, the size of $B$. We use $10^4$ sampled code words for each value of $\ell$ and $B$. We set $T=5$. The success rate of the algorithm rapidly decays as we require it to decode larger errors.}
    \label{fig:benchmark_gallager_bp}
\end{figure}

For our numerical studies, we have empirically characterized Algorithm \ref{alg:gallager_bp}'s ability to correct multiple errors for multiple random ILP instances of increasing size mapped to max-XORSAT problems. This is shown in Fig. \ref{fig:benchmark_gallager_bp}. Here we see an estimate of the success rate for decoding the zero code-word $\mathbf{c}$ with $\ell$ random bit flips as a function of both, the number of errors and the problem size, which we take as the total number of entries of the parity check matrix $B^T$. We see that the expected efficacy of our belief propagation algorithm significantly degrades as we require it to correct for larger errors. This is both a feature of the hard-threshold nature of the belief propagation algorithm as well as the properties of the LDPC code defined by the matrix $B^T$. Note that the method cannot decode all errors faithfully and thus its quantum implementation will not be able to completely uncompute the syndrome.

\begin{algorithm}[h]
\caption{Binary Belief Propagation (BP1) }
\begin{algorithmic}[1]
\STATE \textbf{Input:} Parity check matrix $B^T \in \{0,1\}^{n \times m}$, code word $y \in \{0,1\}^m$, maximum iterations $T$
\STATE \textbf{Output:} Decoding success flag \texttt{is\_feasible}, decoded bits vector \texttt{bits}

\STATE Initialize \texttt{bits} $\gets y$
\STATE $n \gets$ number of variable nodes (columns of $B^T$)

\FOR{$t = 0$ to $T-1$}
    \STATE $\texttt{syndrome} \gets (B^T \cdot \texttt{bits}) \bmod 2$
    \IF{all entries of \texttt{syndrome} are zero}
        \STATE \textbf{return} (\texttt{True}, \texttt{bits}) \COMMENT{Decoding successful}
    \ENDIF
    \STATE Initialize \texttt{flip\_candidates} as zero vector of same shape as \texttt{bits}
    \FOR{$j = 0$ to $n-1$}
        \STATE $\texttt{connected\_cnodes} \gets \{i \mid B^T_{i,j} = 1\}$
        \STATE $\texttt{threshold} \gets |\texttt{connected\_cnodes}|$
        \STATE $\texttt{unsatisfied} \gets$ number of $i \in \texttt{connected\_cnodes}$ with $\texttt{syndrome}[i] = 1$
        \IF{$\texttt{unsatisfied} \geq \texttt{threshold}$}
            \STATE $\texttt{flip\_candidates}[j] \gets 1$
        \ENDIF
    \ENDFOR
    \STATE $\texttt{bits} \gets (\texttt{bits} + \texttt{flip\_candidates}) \bmod 2$
\ENDFOR

\IF{all entries of \texttt{bits} are zero}
    \STATE \textbf{return} (\texttt{True}, \texttt{bits})
\ELSE
    \STATE \textbf{return} (\texttt{False}, \texttt{bits})
\ENDIF
\end{algorithmic}\label{alg:gallager_bp}
\end{algorithm}

\subsection{Quantum circuit implementation}\label{sec:quantum_circuit}

\subsubsection{Preliminaries}\label{sec:preliminary_quantum_circuit_blocks} 
In this section, we provide a detailed description of a quantum circuit to implement the DQI algorithm for a given max-XORSAT problem defined by  a parity check matrix $B^T$ and constraint vector $\mathbf{v}$. In our implementation, we propose a quantum-coherent version of the BP1 algorithm~\cite{10.5555/1795974,PhysRevLett.84.1355}, detailed in Algorithm~\ref{alg:gallager_bp}. The proposed quantum circuit for the decoding step constitutes one of the primary contributions of this work, and represents a key difference from the DQI quantum circuit proposed in Ref.~\cite{patamawisut2025quantumcircuitdesigndecoded}. There, two quantum circuits for the decoding step are implemented: a lookup table decoder and a Gauss-Jordan elimination decoder. The lookup table method, while more general, requires classically precomputing every possible error pattern and its corresponding syndrome. This approach becomes inefficient as the system size increases due to exponential growth in the number of possible patterns.  The Gauss-Jordan decoder is guaranteed to succeed when $B$ is a square and full-rank matrix, i.e. when $B$ defines a uniquely solvable linear system. In the general case, when $B$ is rectangular, the Gauss-Jordan algorithm will find valid code words, however, in general, it will fail to find the correct one, i.e., the one that requires the least bit flips to be correct and brings the corrupted word to the all zero message. In Appendix~\ref{app: GJapp} we show that the success rate of the Gauss-Jordan decoder for the non-square parity-check matrices tested in Fig.~\ref{fig:benchmark_gallager_bp} is clearly lower than the success rate achieved with the BP1 decoder we use in our quantum circuit implementation. To the best of our knowledge, our work presents the first quantum algorithm to implement belief-propagation decoding. In fact, our implementation works for non-square matrices with high probability. 

The whole DQI quantum circuit, based on the algorithm proposed in Ref.~\cite{jordan2024optimizationdecodedquantuminterferometry}, consists of the following steps:
\begin{itemize}
    \item \textbf{Dicke State Preparation:} \\
    To prepare the quantum state in Eq.~\eqref{eq: Dicke state}, we adopt the quantum circuit construction from Ref.~\cite{patamawisut2025quantumcircuitdesigndecoded}. The implementation is based on two steps, a Unary Amplitude Encoding~\cite{HoyerSpalek2005} step to first prepare the state 
\begin{equation}
    \sum_{k=0}^\ell w_k|k\rangle
\end{equation}
and the following conversion of the above state into the desired superposition of Dicke states~\cite{Bartschi_2022}. We refer to Ref.~\cite{patamawisut2025quantumcircuitdesigndecoded} for implementation details and a comprehensive analysis of the required quantum resources involved in this initial step. 
\item \textbf{Encoding phases:}

To prepare the state in Eq.~(\ref{eq: encoding phases}), we simply apply $Z$ gates to selected qubits, as succinctly described  in Algorithm~\ref{alg:apply_z_gates}. 

\begin{algorithm}[H]
\caption{Apply $Z$ gates based on vector $v$}
\begin{algorithmic}[1]
\FOR{$i = 0$ to $m - 1$}
    \IF{$v[i] =1$}
        \STATE Apply $Z$ gate on qubit $y[i]$
    \ENDIF
\ENDFOR
\end{algorithmic}
\label{alg:apply_z_gates}
\end{algorithm}

\item \textbf{Encoding syndrome:} 

This step can be achieved by implementing a matrix vector multiplication (mod 2) with the standard quantum algorithm described in Algorithm~\ref{alg:syndrome encoding}.

\begin{algorithm}[H]
\caption{Syndrome encoding}
\begin{algorithmic}[1]
\FOR{$i = 0$ to $n-1$} 
    \FOR{$j = 0$ to $m-1$} 
        \IF{$B^T[i, j] = 1$}
            \STATE Apply $\mathrm{CNOT}$ gate with control qubit $\texttt{y}[j]$ and target qubit $\texttt{syndrome}[i]$
        \ENDIF
    \ENDFOR
\ENDFOR
\end{algorithmic}\label{alg:syndrome encoding}
\end{algorithm}

\end{itemize}

\subsubsection{Quantum circuit for the binary belief propagation algorithm}\label{sec:quantum_bf_gallager}

\begin{figure}[H]
    \centering
    \includegraphics[width=0.6\linewidth]{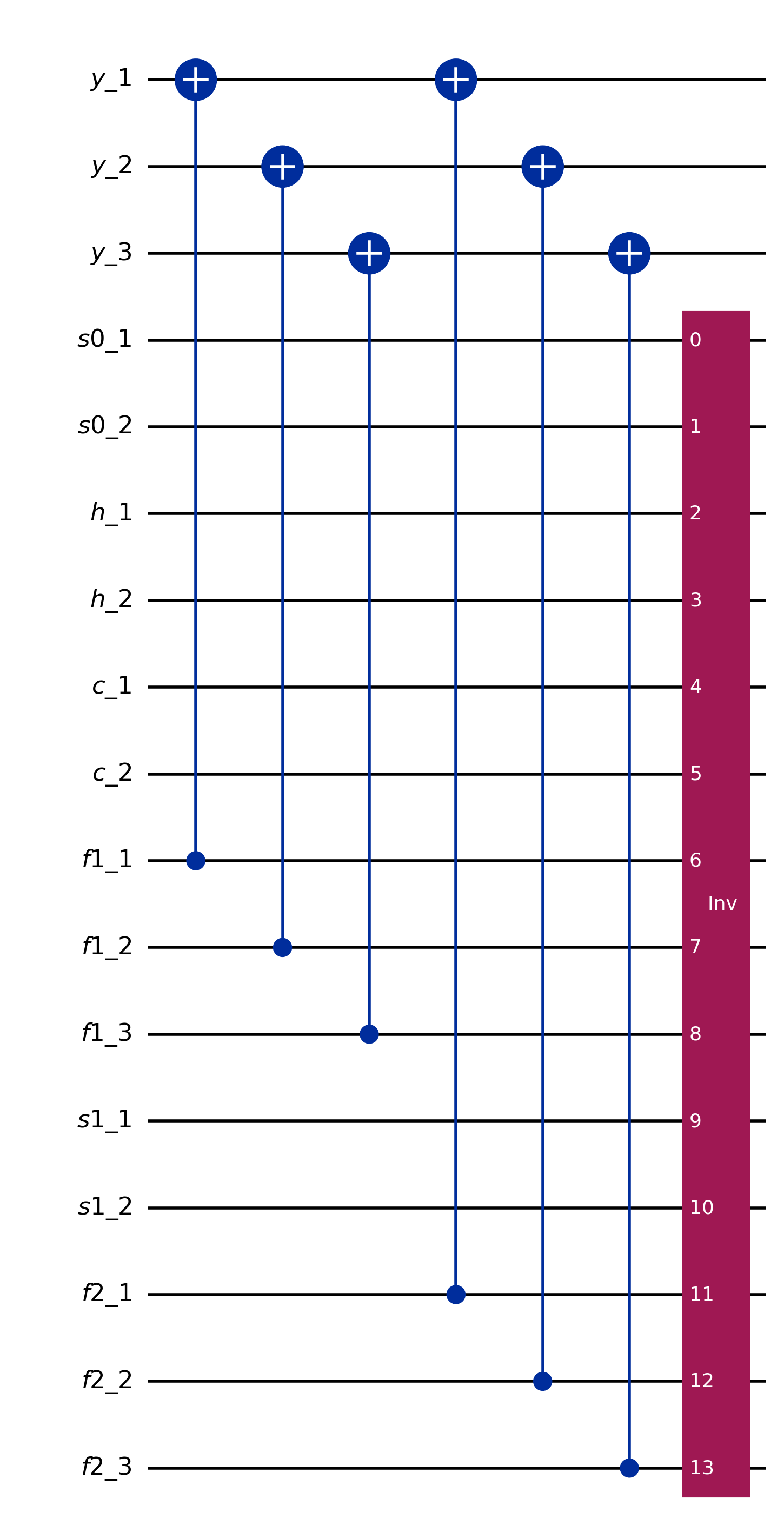}
    \caption{Final step of the quantum‐circuit implementation of the BP1 algorithm for the \(m=3, n=2\) example defined by the parity‐check matrix in Eq.~\ref{eq:matrix small example}. The label \texttt{Inv} indicates the inverse of the entire circuit—excluding the final CNOT gates controlled by the \textit{flip}  registers (\texttt{fi}) on the \textit{message} (\texttt{y}) register.}  
\label{fig:example_quantum_circuit_iteration_3}
\end{figure}

In this section, we provide details of the quantum subroutine we propose for the decoding step of the DQI algorithm. The goal of this step is to uncompute the \emph{message} register, that is:

\begin{equation}
    \sum_{k=0}^{\ell} w_k \frac{1}{\sqrt{\binom{m}{k}}} 
    \sum_{|\mathbf{y}| = k} (-1)^{\mathbf{v} \cdot \mathbf{y}} 
    \left| \mathbf{y} \right\rangle \left| B^T \mathbf{y} \right\rangle 
    \longrightarrow 
    \sum_{k=0}^{\ell} w_k \frac{1}{\sqrt{\binom{m}{k}}} 
    \sum_{|\mathbf{y}| = k} (-1)^{\mathbf{v} \cdot \mathbf{y}} 
    \left| 0^{\otimes m} \right\rangle \left| B^T \mathbf{y} \right\rangle.
\end{equation}
We implement a quantum-coherent version of the BP1 algorithm described in Algorithm~\ref{alg:gallager_bp}. To better illustrate the design of the corresponding quantum circuit and complement the discussion in this section, Figs.~\ref{fig:example_quantum_circuit_iteration_1},~\ref{fig:example_quantum_circuit_iteration_2} and~\ref{fig:example_quantum_circuit_iteration_3} present three sequential sub-circuits representing the BP1 quantum circuit over two iterations ($T=2$), applied to a max-XORSAT instance with \( m = 3 \) constraints and \( n = 2 \) variables. This instance is defined by the matrix 
\begin{equation}
\label{eq:matrix small example}
B =
\begin{bmatrix}
1 & 0 \\
1 & 1 \\
0 & 1
\end{bmatrix}.
\end{equation}

The first step is to extend the quantum circuit originally consisting of the \emph{message} (size $m$ qubits) and \emph{syndrome} registers (size $n$ qubits), denoted by \texttt{y} and \texttt{s0} in Fig.~\ref{fig:example_quantum_circuit_iteration_1}, respectively, by adding ancillary registers. Specifically, we introduce two ancillary registers: \emph{hamming} and \emph{comparator}, denoted as \texttt{h} and \texttt{c} in Fig.~\ref{fig:example_quantum_circuit_iteration_1}, each of size $\left\lceil \log_2(t + 1)\right\rceil$ qubits, where $t$ is the maximum number of ones in any row of the matrix $B$ (i.e., the largest number of variables appearing in any single constraint).

We now describe the algorithm, which proceeds for $T$ iterations, which amount to subsequent repetitions of the following quantum operations. Figs.~\ref{fig:example_quantum_circuit_iteration_1} and~\ref{fig:example_quantum_circuit_iteration_2} illustrate the first and second iterations. For each iteration $i$, we do the following:

\begin{enumerate}
    \item Create a new \emph{flip} register, \texttt{fi} in Fig~\ref{fig:example_quantum_circuit_iteration_1}, of size $m$ qubits.
    \item Iterate over each constraint, i.e., each row of $B$:
    \begin{itemize}
        \item Generate a list of variable indices involved in the current constraint. That is, for each row of $B$, store the column indices where $B[\text{row}, \text{column}] = 1$. This list may contain up to $t\leq n$ elements.

        \item Compute the Hamming weight, $H(\cdot)$, of the relevant subset of qubits in the \emph{syndrome} register \texttt{si'} where \texttt{i'=i-1}, selecting only those qubits indexed by the list above and store the result in the \textit{hamming} register \texttt{h}. To implement a quantum subroutine that coherently computes the Hamming weight of a given set of qubits, we construct a quantum circuit based on the quantum network proposed in Ref.~\cite{Kaye_2001}. The structure of this circuit is shown in Fig.~\ref{fig: hamming circuit}. A key component of the circuit is a unitary gate that performs binary increments, denoted as $U_{+1}$. This gate applies a modular addition of one to a binary-encoded input state:
        \begin{equation}
            U_{+1} \left| \mathrm{bin}(a) \right\rangle = \left| \mathrm{bin}(a + 1) \right\rangle,
        \end{equation}
        for any non-negative integer $a$.
    The circuit implementation to construct the $U_{+1}$ gate is described in Algorithm~\ref{alg:increment_by_one}.

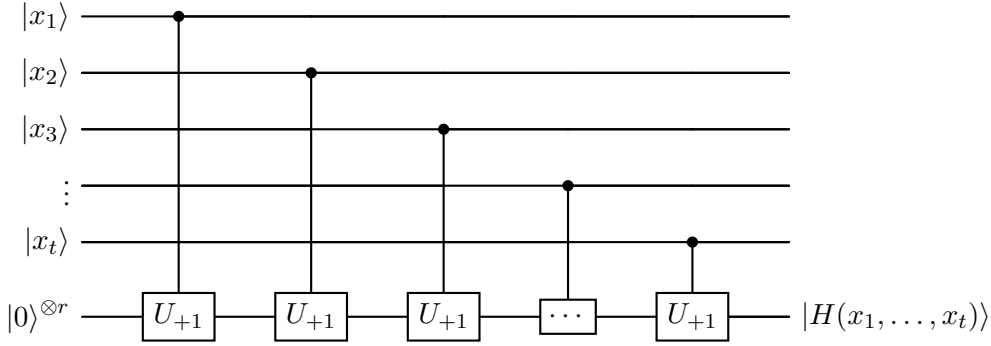
\begin{figure}[ht]
  \centering
  \begin{quantikz}[row sep=0.6cm, column sep=0.8cm]
    \lstick{$\ket{x_1}$} & \ctrl{5} & \qw      & \qw      & \qw      & \qw & \qw \\
    \lstick{$\ket{x_2}$} & \qw      & \ctrl{4} & \qw      & \qw      & \qw& \qw  \\
    \lstick{$\ket{x_3}$} & \qw      & \qw      & \ctrl{3} & \qw      & \qw & \qw \\
    \lstick{$\vdots$}    & \qw      & \qw      & \qw      & \ctrl{2} & \qw& \qw  \\
    \lstick{$\ket{x_t}$} & \qw      & \qw      & \qw      & \qw  &\ctrl{1} & \qw \\
    \lstick{$\ket{0}^{\otimes r}$} 
      & \gate{U_{+1}} 
      & \gate{U_{+1}} 
      & \gate{U_{+1}} 
      & \gate{\cdots} 
      & \gate{U_{+1}}
      & \qw 
      \rstick{$\ket{H(x_1,\dots,x_t)}$} \\
  \end{quantikz}
  \caption{Quantum circuit construction inspired by Ref.~\cite{Kaye_2001} computing $\ket{H(x_1,\dots,x_t)}$ via successive controlled $U_{+1}$ gates. The register where the Hamming weight is stored must be of at least size $r = \left\lceil \log_2(t + 1) \right\rceil$ qubits.}
  \label{fig: hamming circuit}
\end{figure}

\begin{algorithm}
\caption{Circuit construction for the $U_{+1}$ gate}
\label{alg:increment_by_one}
\begin{algorithmic}[1]

\STATE Initialize a quantum circuit with $r$ qubits
\FOR{$i = 0$ to $r-1$}
   
    \IF{$i = 0$}
        \STATE Apply $X$ gate to qubit $0$ \COMMENT{Flip the least significant bit}
    \ELSE
         \STATE Set \texttt{controls} = $\{0, 1, \dots, i-1\}$
        \FOR{each qubit index $c$ in \texttt{controls}}
            \STATE Apply $X$ gate to qubit $c$ \COMMENT{Prepare for multi-controlled Toffoli}
        \ENDFOR
        \STATE Apply multi-controlled $X$ (Toffoli) gate with \texttt{controls} targeting qubit $i$
        \FOR{each qubit index $c$ in \texttt{controls}}
            \STATE Apply $X$ gate to qubit $c$ \COMMENT{Uncompute preparation}
        \ENDFOR
    \ENDIF
\ENDFOR

\end{algorithmic}
\end{algorithm}

        \item Compare the Hamming weight obtained to the number of ones in the constraint (i.e., the length of the index list). If the Hamming weight is greater than or equal to the number of ones, set the first qubit of the \textit{comparator} register to 1. This is achieved using the \texttt{IntegerComparator} gate from Qiskit~\cite{javadiabhari2024quantumcomputingqiskit}.

        \item Apply a CNOT gate controlled on the first qubit of the \textit{comparator} register, targeting the qubit in the \textit{flip} register \texttt{fi\_k} corresponding to the current constraint $k$.

        \item Uncompute the \textit{hamming} and \textit{comparator} registers by applying the inverse of the operations used to compute them. This allows us to reuse these registers for the next constraint and iteration.
    \end{itemize}

    \item At the end of each iteration (except for the last iteration), update the syndrome for use in the next iteration. To preserve the original \textit{syndrome} register (as required by Eq.~\eqref{eq: decoding simple goal}), we introduce a new \textit{syndrome} register, \texttt{si} in Fig.~\ref{fig:example_quantum_circuit_iteration_1}, to represent the updated syndrome after the iteration. For each term in the superposition, this register is expected to be in the state:
    \begin{equation}
    \label{Eq: new_register}
    |B^T \mathbf{y}_i\rangle 
    = |B^T(\mathbf{y}_{i-1} \oplus \mathbf{flip}_i)\rangle 
    = |B^T \mathbf{y}_{i-1} \oplus B^T \mathbf{flip}_i\rangle.
\end{equation}

    We construct the \textit{syndrome} register \texttt{si} in two steps. First, we add (mod 2) the previous syndrome register into the new one using CNOT gates between corresponding qubits. Second, we apply the syndrome encoding subroutine (see Algorithm~\ref{alg:syndrome encoding}) controlled on the \textit{flip} register \texttt{fi} to update the new syndrome according to Eq.~\eqref{Eq: new_register}. 

\end{enumerate}

\begin{figure}[H]
    \centering
    \includegraphics[width=0.92\linewidth]{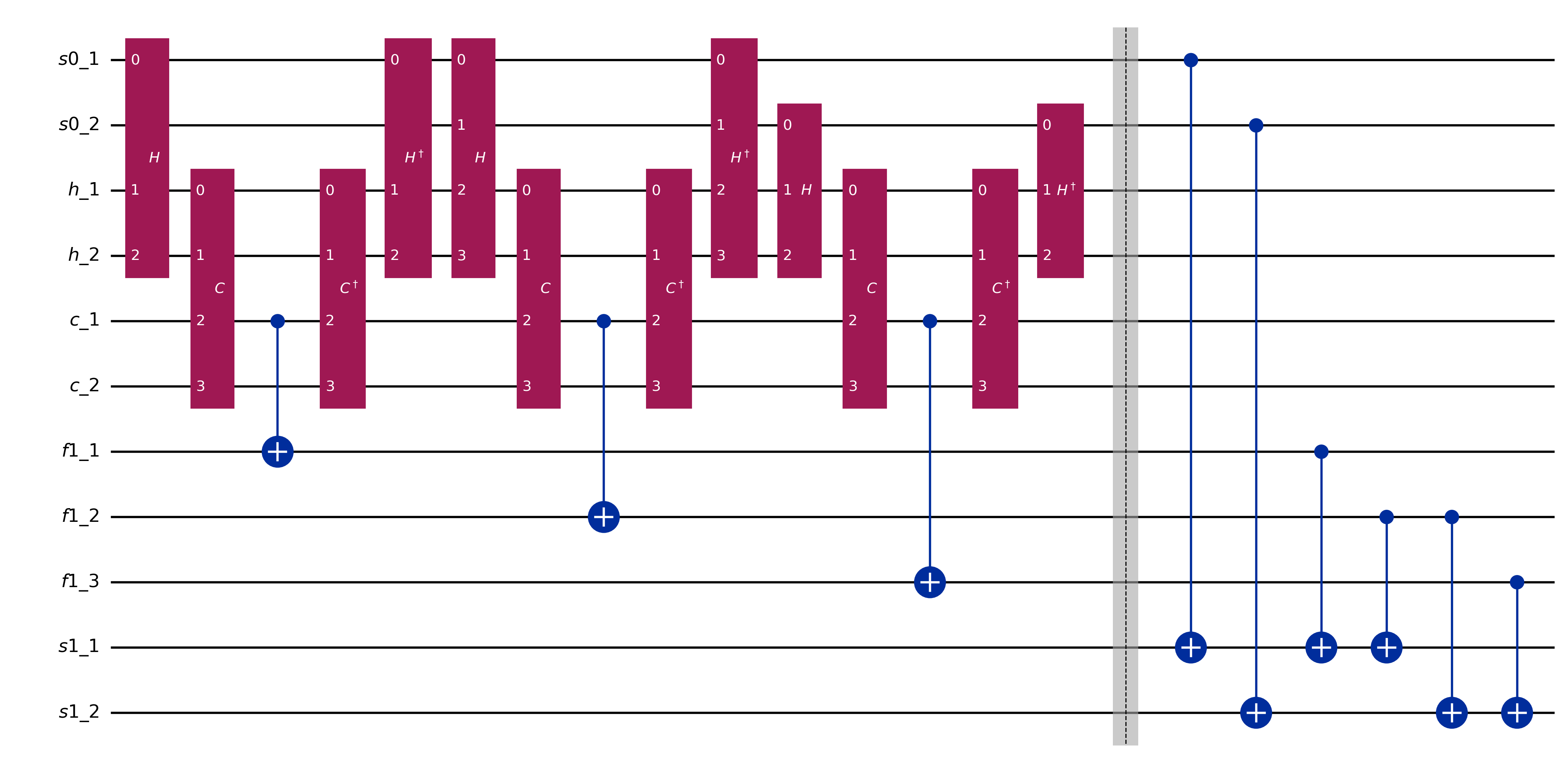}
    \caption{First iteration of the quantum‐circuit implementation of the Binary Belief Propagation algorithm, BP1, for the \(m=3, n=2\) example defined by the parity‐check matrix in Eq.~\ref{eq:matrix small example}.  Here, \(C\) denotes Qiskit’s \texttt{IntegerComparator} subroutine and \(H\) denotes the Hamming‐weight subroutine shown in Fig.~\ref{fig: hamming circuit}. At the right of the barrier, we find the gates corresponding to the subroutine updating the syndrome. }  

    \label{fig:example_quantum_circuit_iteration_1}
\end{figure}

\begin{figure}[H]
    \centering
    \includegraphics[width=0.92\linewidth]{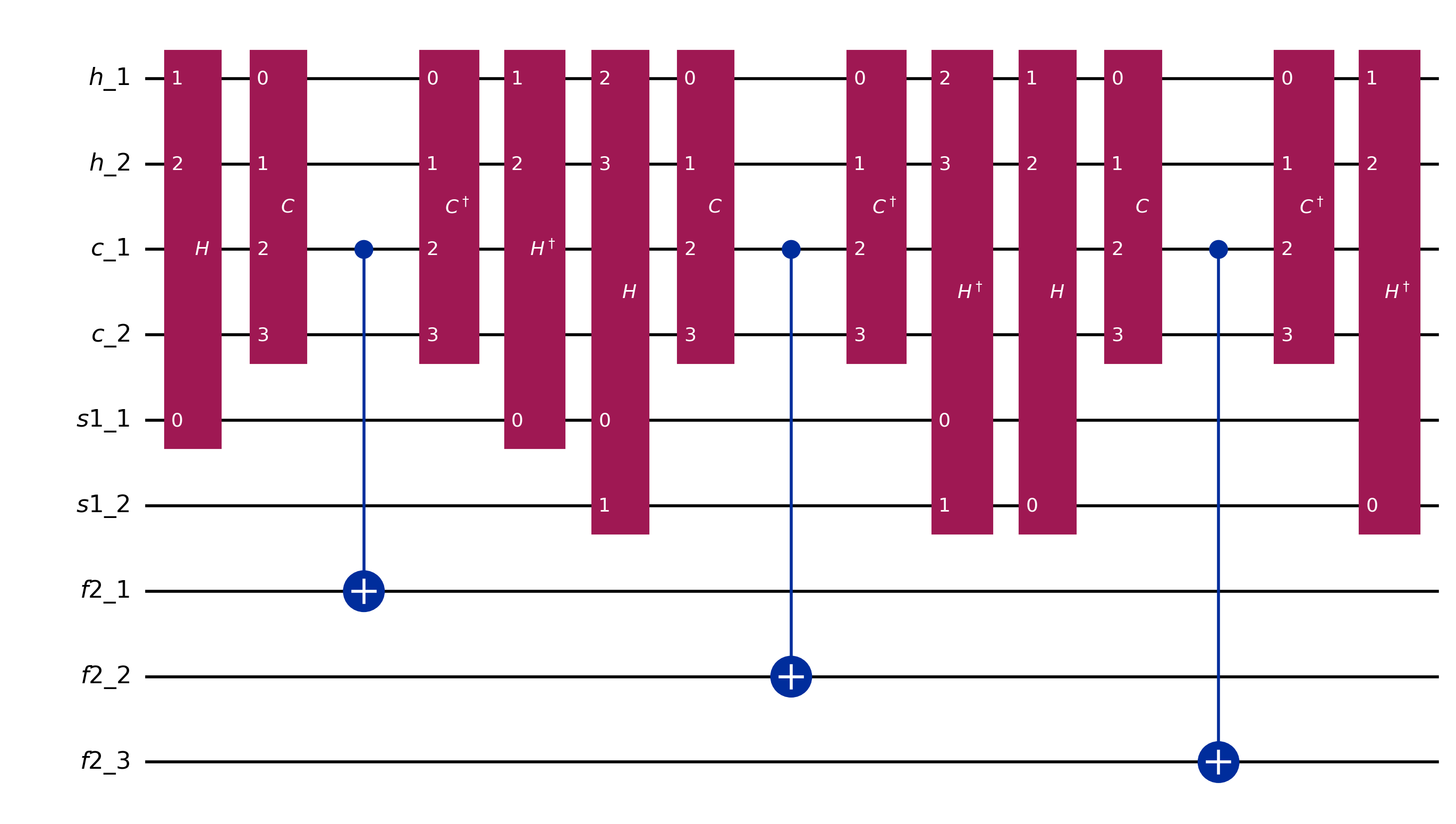}
    \caption{Second iteration of the quantum‐circuit implementation of the Binary Belief Propagation algorithm, BP1, for the \(m=3, n=2\) example defined by the parity‐check matrix in Eq.~\ref{eq:matrix small example}.  Here, \(C\) denotes Qiskit’s \texttt{IntegerComparator} subroutine and \(H\) denotes the Hamming‐weight subroutine shown in Fig.~\ref{fig: hamming circuit}.}  

    \label{fig:example_quantum_circuit_iteration_2}
\end{figure}

Finally, after completing all $T$ iterations, we sum (mod 2) all \textit{flip} registers \texttt{fi} and the original \textit{message} register \texttt{y}. This is implemented by applying a CNOT gate from each qubit in each \textit{flip} register \texttt{fi} to the corresponding qubit in \texttt{y}. This operation transforms the \textit{message} register as follows:

\begin{equation}
    |\mathbf{y}\rangle \longrightarrow |\mathbf{y} \oplus \mathbf{flip}_1 \oplus \mathbf{flip}_2 \oplus \cdots \oplus \mathbf{flip}_T\rangle.
\end{equation}
If the decoder succeeds, this transformation uncomputes the \textit{message} register as desired. This final operation is illustrated in Fig.~\ref{fig:example_quantum_circuit_iteration_3}. 

Before proceeding with the rest of the DQI algorithm, we need to uncompute all ancillary qubits used in the belief propagation decoding. This is done by applying the inverse of all operations carried out before the final modulo-2 addition step. This reversal effectively uncomputes all ancillary registers, while leaving the original \textit{syndrome} and \textit{message} registers intact, since the former is only ever used as a control and the latter is only modified at the final non-uncomputed step.

An important feature of the quantum BP1 implementation is that if the algorithm succeeds in fewer than $T$ iterations, e.g., after $d < T$ iterations, then continuing the remaining iterations will not affect the final result. Specifically, if
\begin{equation}
    |\mathbf{y}\rangle \longrightarrow |\mathbf{y} \oplus \mathbf{flip}_1 \oplus \mathbf{flip}_2 \oplus \cdots \oplus \mathbf{flip}_d\rangle = |0^{\otimes m}\rangle,
\end{equation}
then all subsequent updates beyond iteration $d$ will leave the state unchanged. This is because the corresponding syndrome becomes $|B^T \mathbf{y}_d\rangle = |0^{\otimes n}\rangle$, and any Hamming weight computations involving the \textit{syndrome} register \texttt{sd} will yield zero. As a result, none of the subsequent CNOT gates will activate, and each additional \textit{flip} register \texttt{fi} for $i > d$ will remain in the all-zero state. This property is crucial in quantum implementations of classical decoders. Unlike in classical algorithms—where we can evaluate intermediate results and terminate the decoding loop early—quantum circuits require the full loop to be executed without mid-circuit measurements. This guarantees that unnecessary iterations beyond the point of successful decoding have no negative effect on the already correct outcome.

\section{Formulation of an industrial optimization problem}\label{sec:problem_formulation}

In the automotive industry, original equipment manufacturers (OEM) face a challenge in product configuration and pricing. Rather than offering vehicles with fixed specifications, modern automakers provide base models with numerous optional features, from aesthetic choices like paint colors and interior materials to functional upgrades such as advanced driver assistance systems and premium sound systems. 

To simplify customer choice and increase profitability, manufacturers bundle related options into so-called packages. These packages typically are priced at a discount compared to purchasing each option individually. Packages reduce both the complexity of the decision for the consumer and enable OEMs to strategically sell lower take-rate options by bundling them with popular features.

A key differentiator to a classical bundling problem is that options are always bought together with a car. OEMs thus plan the so-called \emph{take rate} for an option, which is the fraction of vehicles sold with that option included. Our model follows this convention and assumes a fixed total vehicle volume: bundled pricing only shifts option shares and never changes the number of cars sold.

We present a simplified version of this automotive bundling problem that serves the purposes of this paper. While a production-level formulation would include additional complexities such as competitive effects, heterogeneous elasticities, and estimation uncertainty, our simplified model captures the essential combinatorial structure.

\subsection{Optimization objective}
Our objective is to maximize total contribution margin. Let $x_{im}$ be a binary decision variable that indicates whether the option $i$ appears in a package $m$. The contribution margin $C_i(P)$ (the per-unit profit after variable costs) of that package is a function of its price $P$, which we define as $(1-d)\sum_{i \in m} P_i$, where $P_i$ are the prices of individual options and $d$ is a relative discount factor $(0 < d < 1)$. 

The objective function becomes:

\begin{equation}
\max_{x \in \{0,1 \}^{N\times M} } \left[\sum_{m=1}^{M} \sum_{i=1}^{N} T_m \cdot C_i((1-d)P_i) \cdot x_{im} + \sum_{i=1}^{N} T'_{\text{only }i} \cdot C_i(P_i)\right],
\end{equation}
where $T_m$ represents the take rate of package $m$, $C_i((1-d)P_i)$ is an expansion of the package's contribution margin into its constituent option parts, and $T'_{\text{only }i} < T_{\text{only }i}$ represents the reduced standalone take rate for option $i$ after package introduction. This reduction is driven by the migration of people buying only option $i$ to the package. Note that customers either purchase a package containing option $i$ or purchase the option standalone, but not both, preventing double-counting in the objective function.

\subsection{Demand model}
Given a set of $N$ individual options with standalone prices $P_i$ and take-rates $T_i$, customers evaluate packages based on the discounted total price $P$, which is smaller than the individual prices $\sum_i P_i$. 

For simplicity's sake, consider a package consisting of options $\mathcal{O}_1$ and $\mathcal{O}_2$. The total customer volume is split over four groups: neither $\mathcal{O}_1$ nor $\mathcal{O}_2$ was bought; only $\mathcal{O}_1$ was bought; only $\mathcal{O}_2$ was bought; and both options were bought. Given that the combination $\mathcal{O}_1$ and $\mathcal{O}_2$ is offered at a discount, there are three types of relevant customer migrations. Customers who previously\ldots
\begin{itemize}
   \item \ldots purchased only option $\mathcal{O}_1$ may upgrade to a package containing both options;
   \item \ldots purchased only option $\mathcal{O}_2$ may upgrade to a package containing both options;
   \item \ldots purchased neither may now be induced to purchase the discounted package.
\end{itemize}

Assuming constant price elasticity of demand, the change in take rate $T$ is given by the incurred change in price:
\begin{equation}
\frac{T'}{T} = f_\epsilon\left(\frac{P'}{P}\right) = \begin{cases}
1 & \text{if } (P'/P)^\epsilon > 1, \\
(P'/P)^\epsilon & \text{otherwise}
\end{cases}
\end{equation}
where $\epsilon<0$ to capture standard price‐elasticity effects and the capping is required to preserve feasibility under fixed total volume of cars, avoiding the mathematical impossibility that more than 100\% of customers change. The capping ensures mathematical consistency with the fixed vehicle volume assumption, though it creates a non-smooth optimization landscape, which can be handled by passing piecewise-linear functions to the used solver. 

For a two-option package $\{\mathcal{O}_1,\mathcal{O}_2\}$, the package take-rate emerges from customer migrations that we described earlier. Using the more concise notation $T_{1,2}$ for $T_{\mathcal{O}_1,\mathcal{O}_2}$, we find:   

\begin{align}
T'_{1,2} &= T_{1,2} + T_{1 \rightarrow 1,2} + T_{2 \rightarrow 1,2} + T_{\emptyset \rightarrow 1,2} \\
\frac{T_{1 \rightarrow 1,2}}{T_{\text{only }1}} &= \left(\frac{(1-d)(P_1 + P_2)}{P_1}\right)^\epsilon \\
\frac{T_{2 \rightarrow 1,2}}{T_{\text{only }2}} &= \left(\frac{(1-d)(P_1 + P_2)}{P_2}\right)^\epsilon \\
\frac{T_{\emptyset \rightarrow 1,2}}{T_\emptyset} &= \left(\frac{(1-d)(P_1 + P_2)}{\kappa}\right)^\epsilon
\end{align}
where $\kappa$ represents a reference price that determines the propensity of model customers who bought neither $\mathcal{O}_1$ nor $\mathcal{O}_2$. We set $\kappa$ to roughly twice the maximum option price, reflecting the idea that a new bundle at that price is unlikely to attract non-buyers. The standalone take rates $T_{\text{only }1}$, $T_{\text{only }2}$, and $T_\emptyset = 1 - T_1 - T_2 + T_{1,2}$ represent the fraction of customers purchasing exactly those option sets before package introduction.

This model readily generalizes to non-overlapping subsets $S_1$ and $S_2$ of packages.
For example, for a three-option package $\{\mathcal{O}_1,\mathcal{O}_2,\mathcal{O}_3\}$, migrations occur from all seven non-empty subsets: $\{\mathcal{O}_1\}$, $\{\mathcal{O}_2\}$, $\{\mathcal{O}_3\}$, $\{\mathcal{O}_1,\mathcal{O}_2\}$, $\{\mathcal{O}_1,\mathcal{O}_3\}$, $\{\mathcal{O}_2,\mathcal{O}_3\}$, with take rates calculated analogously using the relative price changes. Migration from the empty set is still modeled the same way. 

Moreover, it is important to emphasize that we allow currently existing options to still be bought independently; that is, there is no forced bundling implied. 

Note that this model requires take rates for each exact subset, which can be measured from historical sales data for existing options. For new options without sales history, (co-)take rates must be estimated using methods such as analogies to similar options in the current portfolio; market research and conjoint analysis; statistical models trained on features of existing options; and expert judgment based on competitive benchmarking. For the purposes of this work, we assume that these take rates are given as input to the optimization problem.

\subsection{Constraints}
The optimization must respect several business and technical constraints:

\begin{align}
\sum_{i \in S} x_{im} &= 1 \quad \forall m, \forall S \in \mathcal{S}_{\text{mandatory}} \quad &\text{(Exactly one from mandatory families)} \\
\sum_{i \in S'} x_{im} &\leq 1 \quad \forall m, \forall S' \in \mathcal{S}_{\text{optional}} \quad &\text{(At most one from optional families)} \\
\sum_{i=1}^{N} x_{im} &\leq O \quad \forall m \quad &\text{(Maximum options per package)} \\
x_{im} + x_{jm} &\leq 1 + \mathcal{B}_{ij} \quad \forall i,j, \forall m \quad &\text{(Compatibility restrictions)} \\
x_{jm} &\leq x_{im} \quad \forall m, \forall (i,j) \in \mathcal{D} \quad &\text{(Dependency requirements)} \\
\sum_{i \in m'} x_{im} &\leq |m'| - 1 \quad \forall m, \forall m' \in \mathcal{E} \quad &\text{(Avoid duplicating existing packages)}
\end{align}
where $\mathcal{S}_{\text{mandatory}}$ and $\mathcal{S}_{\text{optional}}$ partition options into families from which exactly 1 option can be chosen, $\mathcal{B}_{ij}$ encodes compatibility of options, $\mathcal{D}$ contains dependency pairs, $\mathcal{E}$ represents existing packages, and $O$ is an integer, indicating the maximum size of a package. Note that $B$ is a symmetric compatibility constraint (option $i$ and $j$ cannot be combined), whereas $\mathcal{D}$ is asymmetric (option $i$ implies $j$, but not vice versa). The use of the sets $\mathcal{S}_{\text{mandatory}}$ and $\mathcal{S}_{\text{optional}}$ allows for a package to be flexibly created: e.g., an interior package \emph{must} have a steering wheel and \emph{must} have a seat type, but \emph{may} have leather inlays. This formulation yields a binary integer program that balances customer preferences with operational constraints to maximize profitability through strategic bundling.

\subsection{Relationship to typical optimization problems}
Our problem represents a hybrid combinatorial optimization problem that unifies covering, partitioning, knapsack, and satisfiability structures of many NP problems \cite{Lucas_2014}. At its core, it exhibits the characteristics of \emph{set cover problems}, where packages (subsets) must meet customer demand while maximizing profit. The mandatory and optional family constraints $\mathcal{S}_{\text{mandatory}}$ and $\mathcal{S}_{\text{optional}}$ create a \emph{partitioning structure}, ensuring non-overlapping selections within packages.

The package size limit $O$ combined with contribution margin maximization resembles a \emph{bounded knapsack variant}, where capacity constraints limit value optimization. Additionally, the compatibility matrix $\mathcal{B}_{ij}$ and dependency set $\mathcal{D}$ introduce logical constraints characteristic of \emph{maximum satisfiability} (MAX-SAT) problems.

\section{Transformation from 0-1 ILP to LDPC code}
\label{sec: trasnformation to max-XOR-Sat}

A \emph{0--1 integer linear program} (0--1 ILP) seeks an assignment of \(n\) binary decision variables \(\mathbf{x}\in\{0,1\}^{n}\) that optimizes a linear objective subject to a finite set of linear (in)equalities.  Let \(\mathbf{c}\in\mathbb{R}^{n}\) be the cost vector, \(A\in\mathbb{R}^{m\times n}\) and \(\mathbf{b}\in\mathbb{R}^{m}\) define \(m\) inequality constraints, and optionally let \(A_{\!=}\in\mathbb{R}^{p\times n}\) and \(\mathbf{b}_{\!=}\in\mathbb{R}^{p}\) specify \(p\) equality constraints.  The general formulation is
\begin{equation}\label{eq:01ILP}
\begin{aligned}
\text{maximize}\quad & \mathbf{c}^{\mathsf T}\mathbf{x} \\
\text{subject to}\quad & A\mathbf{x}\;\le\;\mathbf{b},\\
                      & A_{\!=}\mathbf{x}\;=\;\mathbf{b}_{\!=},\\
                      & \mathbf{x}\in\{0,1\}^{n},
\end{aligned}
\end{equation}
where “\(\le\)” is interpreted component‑wise. Equivalently, a minimization problem is obtained by negating the objective coefficients. The binary integrality requirement, \(\mathbf{x}\in\{0,1\}^{n}\), distinguishes 0--1 ILPs from continuous linear programs and gives rise to NP‑hard combinatorial structure, making them a powerful modeling tool for various industrial combinatorial optimization problems. 

Mapping a 0-1 ILP problem instance to an equivalent instance of max-XORSAT can be divided into two steps: 

\begin{enumerate}
    \item Mapping the constraints into an equivalent max-XORSAT instance, namely $\bf x$ is feasible in \eqref{eq:01ILP} iff its binary representation $\bf s$ maximally satisfies the equation $M{\bf s}={\bf y}$ mod 2 for some matrix $M$ and vector $\bf y$ of size polynomial in $n$. Details of this will be presented in Section \ref{subsec:constraints}.
    \item Reducing the objective function maximization to solving max-XORSAT. Similar to the case of weighted partial max-SAT\footnote{A Weighted Partial max-SAT (WPMS) consists of a set of weighted clauses $\phi=\{(C_1,w_1),(C_2,w_2),\cdots,(C_m,w_m)\}$ where the goal is to find an assignment such that the set of violated (or falsified) clauses have the minimal total cost. Like ILP, a WPMS instance $\phi=\phi_H\cup\phi_S$ can be partitioned into ``hard clauses" $\phi_H$ that have infinite weight, meaning that they are equivalent to constraints in ILP and must be satisfied, and ``soft clauses" $\phi_S$ each of which has a finite weight and may or may not be satisfied.}, where one of the most successful strategy is to iteratively find the optimal solution by using a SAT solver as an oracle on each iteration \cite{Ansotegui2013SATBasedmaxSAT,Morgado2013IterativeCoreGuided}, here we adopt a similar strategy where an ILP instance is reduced to one or more calls to a max-XORSAT solver. Details are presented in Section \ref{subsec:obj}.
\end{enumerate}

An additional concern in reducing ILP to max-XORSAT that is specific to the context of this study is to ensure that the resulting max-XORSAT instance $B{\bf y}={\bf v}\text{ mod 2}$ has its matrix $B$ such that $B^T$ is the parity check matrix of an error correcting code of reasonably high distance. This poses structural demands on $B$. For instance, a simple way of reducing the objective function maximization to a max-XORSAT instance, at least in the case where the vector $\bf c$ consists of non-negative integers, is by repeating equation $x_i=1$ for $c_i$ times for each $i\in\{1,2,\cdots,n\}$ in the max-XORSAT instance. This way, satisfying each $x_i=1$ will incur a reward $c_i$ since there are $c_i$ identical equations. However, such an approach is ruled out (in addition to being inefficient with respect to $\sum_i|c_i|$) because repeated rows in $B$ leads to repeated columns in $B^T$, which means the code with $B^T$ as its parity check matrix has only distance 2, a property that disqualifies it as a viable error correcting code.

\subsection{Encoding pseudo-Boolean constraints} \label{subsec:constraints}

The key building block that supports the reduction from ILP to max-XORSAT is to construct a max-XORSAT instance that embeds a pseudo-Boolean constraint of the form
\begin{equation}
    a^{(1)}x_1+a^{(2)}x_2+\cdots+a^{(n)}x_n \bowtie b
\end{equation}
where each $a^{(k)}$ is a positive integer, $x_k\in\{0,1\}$ for any $k$ from 1 to $n$, $\bowtie$ can be either an equality $=$ or an inequality $\le$, and $b$ is also a positive integer. 

A basic question in mapping a pseudo-Boolean constraint to max-XORSAT is how to represent integer addition in binary operations XOR $\oplus$ and AND $\cdot$. Indeed, for adding two integers $u=u_0u_1\cdots u_{\ell-1}$ and $v=v_0v_1\cdots v_{\ell-1}$, each represented in binary form, schoolbook addition means that at the $k$-th position, $1\le k\le\ell-1$, we have the bits $v_k$, $u_k$ and the carry bit $c_{k-1}$ as input, and the output as the sum bit $s_k$ and the next carry bit $c_k$ as
\begin{equation} \label{eq:add_bits}
\begin{aligned}
    s_k & = v_k + u_k + c_{k-1}\text{ mod 2} \\
    c_k & = (v_k\cdot u_k) + (v_k\cdot c_{k-1}) + (u_k\cdot c_{k-1})\text{ mod 2}.
\end{aligned}
\end{equation}
Here $s_k= v_k + u_k + c_{k-1}\text{ mod 2}$ is already in max-XORSAT form, which leaves the AND operation. 

\paragraph{\emph{AND gadget.}}
Consider the following max-XORSAT instance for $x,y,z\in\{0,1\}$
\begin{equation}\label{eq:and}
    \left\{
\begin{aligned}
  x+y+z=1\text{ mod 2} \\
  x+z=0 \text{ mod 2}\\
  y+z=0\text{ mod 2} \\
  z=0\text{ mod 2}.
\end{aligned}
\right.
\end{equation}
Then the above set of equations is maximally satisfied (3 out of 4) iff $z=x\cdot y$. 

\paragraph{\emph{CARRY gadget.}} The carry bit $c_k$ in Equation \eqref{eq:add_bits} can be calculated by introducing auxiliary variables $p_k=v_k\cdot u_k$, $q_k=v_k\cdot c_{k-1}$, and $r_k=u_k\cdot c_{k-1}$, leading to $c_k = p_k+q_k+r_k\text{ mod 2}$. The full max-XORSAT instance that embeds \eqref{eq:add_bits} can then be expressed as (all equations in mod 2)
\begin{equation}\label{eq:carry_gadget}
    \left\{
\begin{aligned}
    c_k + p_k + q_k + r_k = 0 \\
    p_k + v_k + u_k = 0 \\
    p_k + v_k = 0 \\
    v_k + u_k = 0 \\
    p_k = 0 \\
    q_k + v_k + c_{k-1} = 0 \\
    q_k + v_k = 0 \\
    v_k + c_{k-1} = 0 \\
    q_k = 0 \\
    r_k + u_k + c_{k-1} = 0 \\
    r_k + u_k = 0 \\
    u_k + c_{k-1} = 0 \\
    r_k = 0 \\
    s_k + u_k + v_k + c_{k-1} = 0.
\end{aligned}
    \right.
\end{equation}
The max-XORSAT instance in \eqref{eq:carry_gadget} has 14 equations, where an assignment maximally satisfies the instance (with 11 out of the 14 equations satisfied in total) if and only if Equation \eqref{eq:add_bits} holds for the bits $s_k$, $v_k$, $u_k$, $c_k$ and $c_{k-1}$.

\paragraph{\emph{CARRY1 gadget.}} A special case of \eqref{eq:add_bits} is the first step of addition (namely $k=0$) which entails
\begin{equation}\label{eq:add_bit_0}
    \begin{aligned}
        s_0 = u_0 + v_0 \text{ mod 2}\\
        c_0 = u_0\cdot v_0.
    \end{aligned}
\end{equation}
Accordingly, the max-XORSAT instance that encodes \eqref{eq:add_bit_0} contains only 5 equations, 4 of which are satisfied in the case of maximum satisfaction (all equations in mod 2):
\begin{equation}
\left\{
    \begin{aligned}
        c_0 + u_0 + v_0 = 1 \\
        c_0 + v_0 = 0 \\
        u_0 + v_0 = 0 \\
        v_0 = 0 \\
        s_0 + v_0 + u_0 = 0.
    \end{aligned}
\right.
\end{equation}
The first four equations essentially come from applying the AND gadget \eqref{eq:and} with $x=c_0$, $y=u_0$ and $z=v_0$.

\begin{figure}
    \centering
    \includegraphics[width=0.3\linewidth]{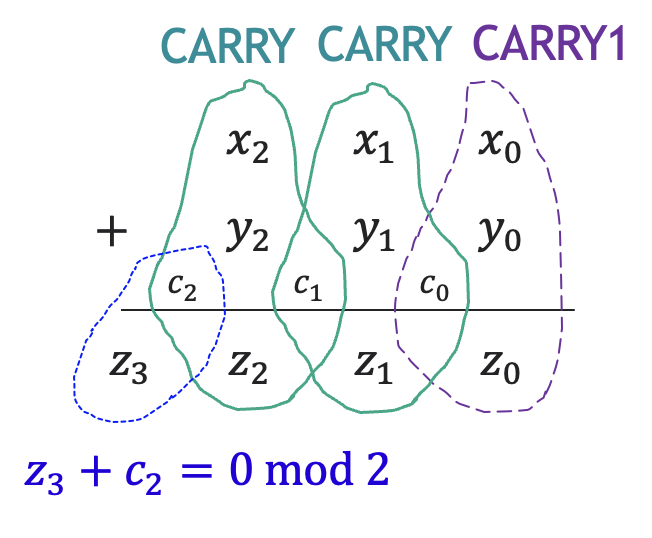}
    \caption{Schematic for building an encoding for integer addition (an ``integer adder circuit") using CARRY and CARRY1 gadgets, using addition of two 3-bit integers $x=x_0x_1x_2$ and $y=y_0y_1y_2$ as an example. }
    \label{fig:ia}
\end{figure}

\paragraph{\emph{Integer adder (IA) circuit.}} The binary arithmetics of adding two integers $u=u_0u_1\cdots u_{\ell-1}$ and $v=v_0v_1\cdots v_{\ell-1}$ to produce a sum $s=s_0s_1\cdots s_{\ell-1}s_\ell$ can be entirely captured by \eqref{eq:add_bit_0} for the initial step and \eqref{eq:add_bits} for addition at positions $1$ through $\ell-1$, combined with the final overflow bit $s_\ell=c_{\ell-1}$. Accordingly, one can encode integer addition into max-XORSAT by applying the CARRY1 gadget to \eqref{eq:add_bit_0} and CARRY gadget(s) to \eqref{eq:add_bits} to yield linear equations mod 2 with binary coefficients. The process is completed by adding the final equation $s_\ell + c_{\ell-1} = 0\text{ mod 2}$ to capture the bit overflow. An example is illustrated in Figure \ref{fig:ia}. For adding general $\ell$-bit integers, with $(\ell-1)$ CARRY gadgets, one CARRY1 gadget and one equation for overflow into $s_\ell$, we have in total 
\begin{equation}\label{eq:xi_ia}
    \xi_\ell^\text{IA}=14(\ell-1)+5+1=14\ell-8
\end{equation}
equations and the maximum number of equations that can be satisfied is 
\[
\eta_\ell^\text{IA}=11(\ell-1)+4+1=11\ell-6.
\]

\begin{figure}
    \centering
    \includegraphics[width=0.65\linewidth]{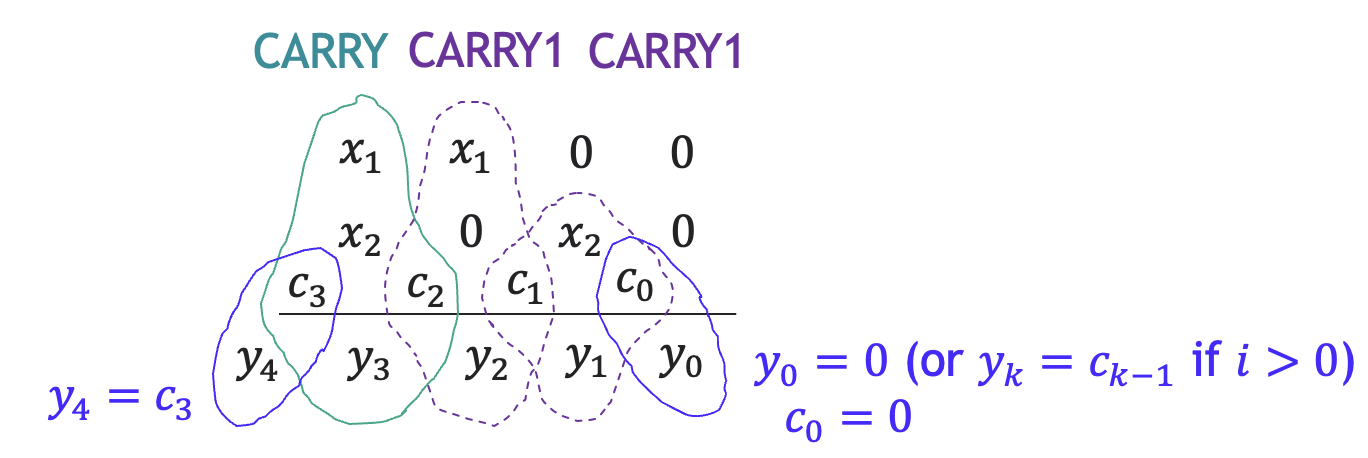}
    \caption{Schematic for building an encoding for weighted integer addition (a ``weighted integer adder") using the constructions in Table \ref{table:wia}, using an example with $a^{(1)}=0011=12$ and $a^{(2)}=0101=10$. Here the four columns are set up to demonstrate the four possible cases (rows in Table \ref{table:wia}) of $a^{(1)}_k$ and $a^{(2)}_k$.}
    \label{fig:wia}
\end{figure}

\paragraph{\emph{Weighted integer adder (WIA).}} As a step closer to pseudo-Boolean constraint encoding, it is of interest to embed weighted addition of two bits in the form of
\begin{equation}\label{eq:ax0ax1}
    a^{(1)}x_1+a^{(2)}x_2 = b
\end{equation}
where $a^{(1)}$ and $a^{(2)}$ are $\ell$-bit integers, $x_1,x_2\in\{0,1\}$, and $b$ is an $(\ell+1)$-bit integer. Here the integers $a^{(1)}$, $a^{(2)}$ and $b$ are given and the goal is to generate a max-XORSAT instance such that a solution satisfies the maximum number of equations iff the bits $x_0$ and $x_1$ satisfy \eqref{eq:ax0ax1}. Similar to the case of the integer adder circuit, we discuss two separate cases: the initial step with $k=0$ and addition at subsequent steps with $1\le k\le\ell-1$. For each case, there are four scenarios corresponding to four different combinations of the values of the $k$-the bit of $a^{(1)}$ and $a^{(2)}$: 00, 01, 10, and 11. So we can tabulate the equation to generate (and gadget to apply if any) in each of the $2\times 4$ situations as in Table \ref{table:wia}. An example is illustrated in Figure \ref{fig:wia}. Let $f_k(u,v)$ be the number of equations generated from adding two bits $u$ and $v$ at position $k$ in the addition process, where $k=0,\cdots,\ell-1$. Also, let $g_k(u,v)$ be the maximum number of equations that can be satisfied for the same definitions of $u$, $v$, and $k$. The values of $f_k$ and $g_k$ can be tabulated as the following:
\[
\begin{tabular}{|c|c|c|c|c|}
    \hline
   $a_k^{(1)}$ & $a_k^{(2)}$ & \text{Range of $k$} & $f_k(a_k^{(1)},a_k^{(2)})$ & $g_k(a_k^{(1)},a_k^{(2)})$ \\
   \hline
   0 & 0 & $k=0$ & 2 & 2 \\
   \hline
   0 & 0 & $k>0$ & 2 & 2 \\
   \hline
   0 & 1 & $k=0$ & 2 & 2 \\
   \hline
   0 & 1 & $k>0$ & 5 & 4 \\
   \hline
   1 & 0 & $k=0$ & 2 & 2 \\
   \hline
   1 & 0 & $k>0$ & 5 & 4 \\
   \hline
   1 & 1 & $k=0$ & 5 & 4 \\
   \hline
   1 & 1 & $k>0$ & 14 & 11 \\
   \hline
\end{tabular}
\]
If we include the final equation $s_\ell=c_{\ell-1}$ or equivalently $s_\ell+c_{\ell-1}=0\text{ mod 2}$, the weighted integer adder embedded in a max-XORSAT instance has 
\begin{equation}\label{eq:xi_wia}
\xi_\ell^\text{WIA}=1+\sum_{k=0}^{\ell-1}f_k(a_k^{(1)},a_k^{(2)})
\end{equation}
equations and the maximum number of equations satisfied is 
\[
\eta_\ell^\text{WIA}=1+\sum_{k=0}^{\ell-1}g_k(a_k^{(1)},a_k^{(2)}).
\]

\begin{sidewaystable}
\centering
\begin{tabular}{|c|l|l|}
\hline
&
$k=0$ &
$1\le k\le\ell-1$ \\
\hline
$\left\{
\begin{aligned}
    a^{(1)}=0 \\
    a^{(2)}=0
\end{aligned}
\right.$
&
$\begin{array}{rcl}
  & 0 & \\
  & 0 & \\[-1.5ex]
c_0 &     & \\
\hline
  & 0 &
\end{array}
\implies
\left\{
\begin{aligned}
    c_0=0 \\
    y_0=0
\end{aligned}
\right.$
& 
$\begin{array}{rcl}
  & 0 & \\
  & 0 & \\[-1.5ex]
c_k & & c_{k-1} \\
\hline
  & 0 &
\end{array}
\implies
\left\{
\begin{aligned}
    c_k=0 \\
    y_k=c_{k-1}
\end{aligned}
\right.$
\\
\hline
$\left\{
\begin{aligned}
    a^{(1)}=1 \\
    a^{(2)}=0
\end{aligned}
\right.$
& 
$\begin{array}{rcl}
  & x_0 & \\
  & 0 & \\[-1.5ex]
c_0 &     & \\
\hline
  & y_0 &
\end{array}
\implies
\left\{
\begin{aligned}
    &c_0=0 \\
    &y_0=x_0
\end{aligned}
\right.$
& 
$\begin{array}{rcl}
  & x_0 & \\
  & 0 & \\[-1.5ex]
c_k & & c_{k-1} \\
\hline
  & y_k &
\end{array}
\implies
\left\{
\begin{aligned}
    &c_k=x_0\cdot c_{k-1} \\
    &y_k=x_0+c_{k-1}
\end{aligned}
\right.$ (CARRY1)
\\
\hline
$\left\{
\begin{aligned}
    a^{(1)}=0 \\
    a^{(2)}=1
\end{aligned}
\right.$
&
$\begin{array}{rcl}
  & 0 & \\
  & x_1 & \\[-1.5ex]
c_0 &     & \\
\hline
  & y_0 &
\end{array}
\implies
\left\{
\begin{aligned}
    &c_0=0 \\
    &y_0=x_1
\end{aligned}
\right.$
& 
$\begin{array}{rcl}
  & 0 & \\
  & x_1 & \\[-1.5ex]
c_k & & c_{k-1} \\
\hline
  & y_k &
\end{array}
\implies
\left\{
\begin{aligned}
    &c_k=x_1\cdot c_{k-1} \\
    &y_k=x_1+c_{k-1}
\end{aligned}
\right.$ (CARRY1)
\\
\hline
$\left\{
\begin{aligned}
    a^{(1)}=1 \\
    a^{(2)}=1
\end{aligned}
\right.$
& 
$\begin{array}{rcl}
  & x_0 & \\
  & x_1 & \\[-1.5ex]
c_0 &     & \\
\hline
  & y_0 &
\end{array}
\implies
\left\{
\begin{aligned}
    &c_0=x_0\cdot x_1 \\
    &y_0=x_0+x_1
\end{aligned}
\right.$ (CARRY1)
& 
$\begin{array}{rcl}
  & x_0 & \\
  & x_1 & \\[-1.5ex]
c_k & & c_{k-1} \\
\hline
  & y_k &
\end{array}
\implies
\left\{
\begin{aligned}
    &c_k=x_0\cdot c_{k-1}+x_1\cdot c_{k-1}+x_0\cdot x_1 \\
    &y_k=x_0+x_1+c_{k-1}
\end{aligned}
\right.$ (CARRY)
\\
\hline
\end{tabular}
\caption{Equations generated for each situation (value of $k$ versus the bit value combination of $a_k^{(1)}$ and $a_k^{(2)}$). Some of the equations are then further mapped to max-XORSAT by the gadgets presented in Section \ref{subsec:constraints}.}
\label{table:wia}
\end{sidewaystable}

\paragraph{\emph{Half weighted adder (HWA).}} As will become clear later, it is useful to embed constraints of the form
\begin{equation}
    ax=y
\end{equation}
where $a=a_0a_1\cdots a_{\ell-1}$ is a given $\ell$-bit integer, and both $x\in\{0,1\}$ and $y=y_0y_1\cdots y_{\ell-1}$ are variables to be determined. Here, $y$ is an (unknown) integer of $\ell$ bits. The equations for max-XORSAT are then for any $k$ from 0 to $\ell-1$, 
\begin{equation}
    \begin{aligned}
        x_k+y_k=0\text{ mod 2,}\quad & \text{if }a_k=1 \\
        y_k=0\text{ mod 2,}\quad & \text{if }a_k=0.
    \end{aligned}
\end{equation}
In total, there are $\xi_\ell^\text{HWA}=\ell$ equations, and an optimal solution should satisfy all $\eta_\ell^\text{HWA}=\ell$ equations.

\begin{figure}
    \centering
    \includegraphics[width=0.75\linewidth]{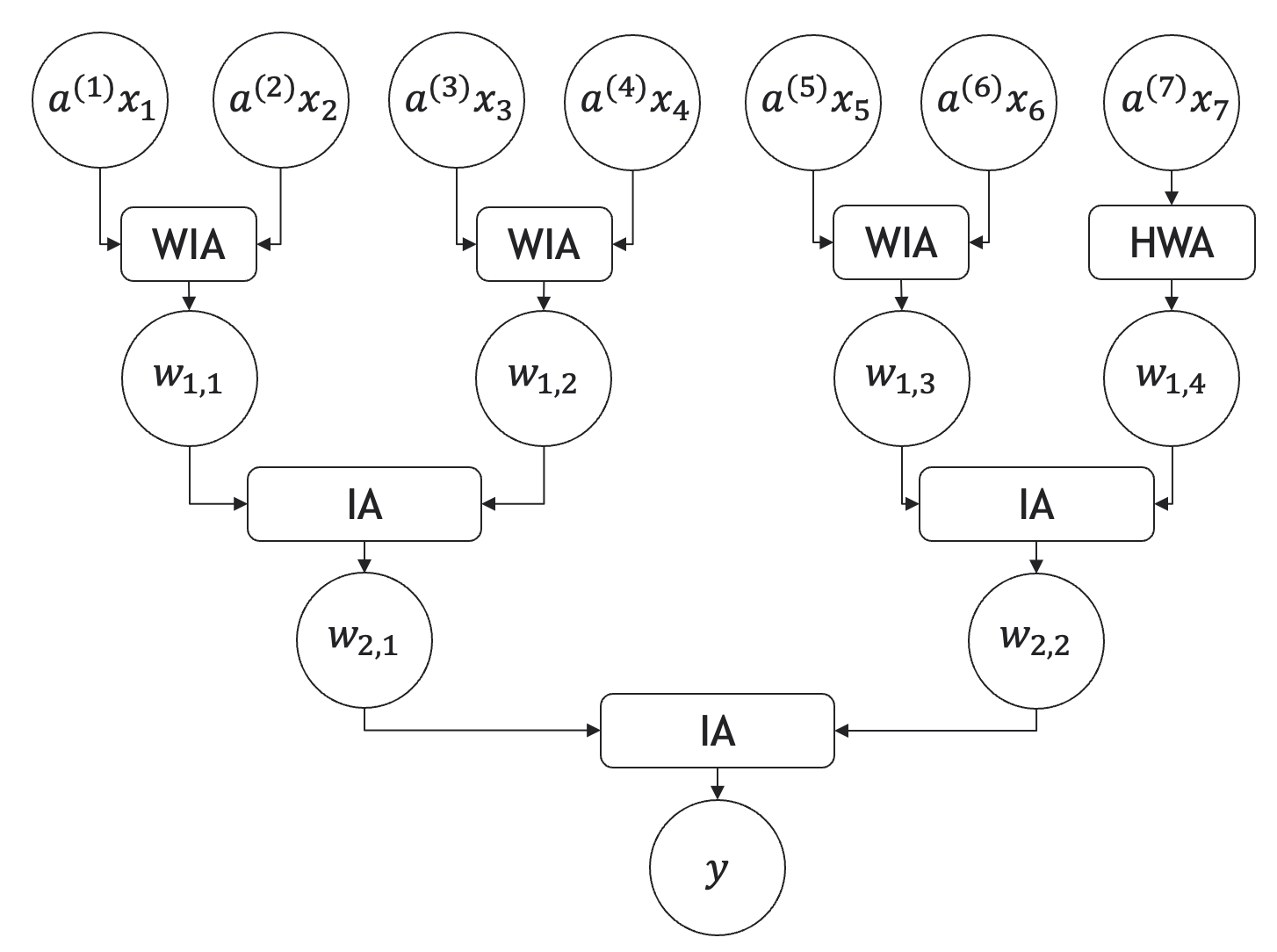}
    \caption{Schematic illustration for the multiple integer adder circuit using the weighted integer adder (WIA), half weighted adder (HWA) and the integer adder (IA) circuits, using as an example the case where there are in total $n=7$ terms. Here $n$ is chosen as an odd number to illustrate how an orphan node at a given level is handled.}
    \label{fig:mia-example}
\end{figure}

\paragraph{\emph{Multiple integer adder (MIA).}} The constructions introduced so far enable embedding weighted addition of multiple bits in the form of
\begin{equation}
    a^{(1)}x_1+a^{(2)}x_2+\cdots+a^{(n)}x_n=y
\end{equation}
where each $a^{(k)}$ is an integer of $\alpha_k$ bits that is given and each bit $x_k\in\{0,1\}$ is a variable to be determined. The output $y$ is an integer of $\ell+J(n)$ bits where the function $J(n)$ for a given $n$ is a value $J$ such that the sequence
$
m_1=\lceil n/2\rceil, \quad
m_{j+1}=\lceil m_j/2\rceil 
$
has $m_J=2$. In other words, $J(n)$ is the number of layers in the binary tree with $n$ leaf nodes (each of which is a term in the sum) constructed using the adders introduced previously, excluding the root node $y$. Figure \ref{fig:mia-example} shows an example of how such tree is constructed. The general scheme for constructing the tree is the following. 
\begin{enumerate}
    \item For the first layer of tree nodes, which consists of WIA(s) and if $n$ is odd, one HWA, amounting to in total $m_1=\lceil n/2\rceil$ terms in the form of $y_k=a^{(2k-1)}x_{2k-1}+a^{(2k)}x_{2k}$ or $y_k=a^{(2k-1)}x_{2k-1}$ for $k=1,\cdots,m_1-1$. For each $k$, each weighted integer adder (WIA) takes as input $a^{(2k-1)}x_{2k-1}$ and $a^{(2k)}x_{2k}$, each containing $\alpha_{2k-1}$ and $\alpha_{2k}$ bits, and produces an output of first layer nodes $w_{1,k}$ of $\max\{\alpha_{2k-1},\alpha_{2k}\}+1$ bits. In the special case of $n=2$, only a single WIA suffices to compute the output $y$ without needing to introduce intermediate $w_{j,k}$ nodes.
    \item For any subsequent layer $j=2,...,J(n)$ with $m_j$ nodes, unless $m_j=2$, we use integer adders (IA) to pair up $w_{j,2k-1}$ and $w_{j,2k}$ nodes and generate $w_{j+1,k}$ for $k=1,...,m_{j+1}-1$ in the next layer. If $m_j$ is odd, there will be an orphan $w_{j,m_j}$ that will be added to the next layer as well, namely $w_{j+1,m_{j+1}}$ and $w_{j,m_j}$ are the same node. If $m_j=2$ then a single IA is used for computing the output $y$ by summing $w_{j,1}$ and $w_{j,2}$.
\end{enumerate}

The total number of equations that can be satisfied in the max-XORSAT instance created from the MIA construction is given as
\begin{equation}\label{eq:mia_tot_eqs}
    \xi_{\alpha_1,\alpha_2,\cdots,\alpha_n}^\text{MIA}=\sum_{k=1}^{\lfloor n/2\rfloor}\xi_{\mu_{1,k}(\alpha_1,\alpha_2,\cdots,\alpha_n)}^\text{WIA}+(n\mod 2)\xi_{\alpha_n}^\text{HWA}+\sum_{j=2}^{J(n)+1}\sum_{k=1}^{m_j}\xi_{\mu_{j,k}(\alpha_1,\alpha_2,\cdots,\alpha_n)}^\text{IA}.
\end{equation}
Similarly, the maximum number of equations that can be satisfied in the max-XORSAT instance created from the MIA construction is given as
\begin{equation}\label{eq:mia_max_eqs}
    \eta_{\alpha_1,\alpha_2,\cdots,\alpha_n}^\text{MIA}=\sum_{k=1}^{\lfloor n/2\rfloor}\eta_{\mu_{1,k}(\alpha_1,\alpha_2,\cdots,\alpha_n)}^\text{WIA}+(n\mod 2)\eta_{\alpha_n}^\text{HWA}+\sum_{j=2}^{J(n)+1}\sum_{k=1}^{m_j}\eta_{\mu_{j,k}(\alpha_1,\alpha_2,\cdots,\alpha_n)}^\text{IA}.
\end{equation}
Here $\mu_{j,k}$ is defined as the number of bits in the variable $w_{j,k}$, with $w_{J(n)+1,1}=y$ being the output variable:
\begin{equation}\label{eq:mu_jk}
    \mu_{j,k}(\alpha_1,\alpha_2,\cdots,\alpha_n)=\left\{
    \begin{array}{ll}
        \max\{\alpha_{2k-1},\alpha_{2k}\}+1, & j=1,\text{ $k<m_1$ or $n$ is even} \\
        \alpha_n, & j=1,\text{ $k=m_1$ and $n$ is odd} \\
        \max\{\mu_{j-1,2k-1},\mu_{j,2k-1}\}+1, & 1<j\le J(n),\text{ $k<m_j$ or $m_{j-1}$ is even} \\
        \mu_{j-1,m_{j-1}}, & 1<j\le J(n),\text{ $k=m_j$ and $m_{j-1}$ is odd.}
    \end{array}
    \right.
\end{equation}
The expressions in \eqref{eq:mia_tot_eqs} and \eqref{eq:mia_max_eqs} come from accounting for the contributions from the different types of gadgets:
\begin{enumerate}
    \item \emph{WIA contribution.} There are in total $\lfloor n/2\rfloor$ WIAs. For each $k=1,2,...,\lfloor n/2\rfloor$, the $k$-th WIA adds $a^{(2k-1)}x_{2k-1}$ and $a^{(k)}x_{2k}$, and produces an output $w_{1,k}$ of $\mu_{1,k}=\max\{\alpha_{2k-1},\alpha_{2k}\}+1$ bits.
    \item \emph{HWA contribution.} One HWA will be needed if and only if $n$ is odd to map the $n$-th term $a^{(n)}x_n$ to an intermediate integer variable.
    \item \emph{IA contribution.} The binary tree of IAa contains in total $J(n)$ layers, with the $j$-th layer containing $m_j$ nodes. Each node is associated with an IA that embeds the addition of the parent nodes. The total number of equations needed for the IA is determined by the number of bits in the output integer, which is $\mu_{j,k}$ for the $k$-th node at the $j$-th layer as described in \eqref{eq:mu_jk}.
\end{enumerate}

\paragraph{\emph{Integer comparator.}} The goal of this gadget is to encode inequality constraints of the form $x<b$ for an unknown integer $x$ of $\ell$ bits and an $\ell$-bit integer $b$ that is known. In this case, comparing the size of the two integers amounts to determining the sign of the difference $x-b$. This is equivalent to computing the quantity
\begin{equation}\label{eq:z_comparator}
    z = x+\bar{b}+1,\text{ where $\bar{b}=2^\ell-b-1$ is the 2's complement of $b$.}
\end{equation}
Here $z=z_0z_1\cdots z_\ell$ is an integer of $\ell+1$ bits. Note that $z=x-b+2^\ell$, hence to see if $x<b$ is equivalently to see if $z_\ell=0$ or $z_\ell=1$. The addition in \eqref{eq:z_comparator} can be embedded into max-XORSAT with a construction similar to the integer adder (IA), except that one of the operands is known, which impacts the details of which gadgets to use.
With exception for the first step in the addition process that calculates $z$, where $\bar{b}_0=0$ leads to
\begin{equation}
    \left\{
    \begin{aligned}
        c_0+x_0 &=0\text{ mod 2}\\
        x_0+z_0 &=1\text{ mod 2}
    \end{aligned}
    \right.
\end{equation}
and $\bar{b}_0=1$ leads to
\begin{equation}
    \left\{
    \begin{aligned}
        c_0 &=1\text{ mod 2}\\
        x_0+z_0 &=0\text{ mod 2},
    \end{aligned}
    \right.
\end{equation}
for any subsequent step $k=1,\cdots,\ell-1$, 
the gadget used to embed the addition step into max-XORSAT also depends on the value of $\bar{b}_k$:

\begin{equation}\label{eq:z_b0_0}
    \begin{array}{rcl}
  & x_k & \\
  & 0 & \\[-1.5ex]
c_k & & c_{k-1} \\
\hline
  & z_k &
\end{array}
\implies
\left\{
\begin{aligned}
    &c_k=x_k\cdot c_{k-1} \\
    &z_k=x_k+c_{k-1}\text{ mod 2}
\end{aligned}
\right.\text{ (CARRY1)}
\end{equation}

\begin{equation}\label{eq:z_b0_1}
    \begin{array}{rcl}
  & x_k & \\
  & 1 & \\[-1.5ex]
c_k & & c_{k-1} \\
\hline
  & z_k &
\end{array}
\implies
\left\{
\begin{aligned}
    &c_k=x_k\cdot c_{k-1}+x_k+c_{k-1}\text{ mod 2} \\
    &z_k=x_k+1+c_{k-1}\text{ mod 2}
\end{aligned}
\right.\text{ (CARRY2)}
\end{equation}
Here the embedding in \eqref{eq:z_b0_1} is a gadget construction that is different from any variant that is introduced so far. It is labeled CARRY2 for distinction. Finally, an equation $z_\ell=c_{\ell-1}$ is added to complete the construction that leads to the computation of the bit $z_\ell$. At this point, one encounters the critical step that determines the nature of integer comparison or inequality that is to be embedded. For embedding $x<b$, add $z_\ell=0$. For embedding $x\ge b$, add $z_\ell=1$.

Following from the previous definitions, the total number of equations in the integer comparator is
\begin{equation}
    \xi_\ell^{<}=\xi_\ell^\ge=1+\sum_{k=0}^{\ell-1}f_k(1,0)=1+2+5(\ell-1)=5\ell-2,
\end{equation}
among which at most
\begin{equation}
    \eta_\ell^<=\eta_\ell^\ge=1+\sum_{k=0}^{\ell-1}g_k(1,0)=1+2+4(\ell-1)=4\ell-1
\end{equation}
equations can be satisfied. Here the functions $f_k(u,v)$ and $g_k(u,v)$ are evaluated at $u=1$, $v=0$ because the values of these functions at $(1,0)$ match the corresponding number of equations and maximum number of satisfiable equations at each position of addition.

\begin{figure}
    \centering
    \includegraphics[width=0.4\linewidth]{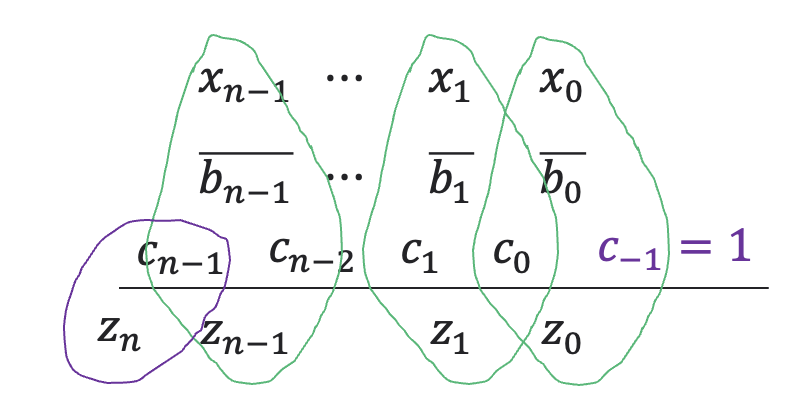}
    \caption{Schematic illustration of how integer comparator computes the intermediate quantity $z=x+\bar{b}+1$. Here we realize the $+1$ factor by introducing an incoming carry bit that is valued as 1.}
    \label{fig:int_comparator}
\end{figure}

\paragraph{\emph{Equality.}} For embedding constraints of the form $x=b$ where $x=x_0x_1\cdots x_{\ell-1}$ is unknown and $b=b_0b_1\cdots b_{\ell-1}$ is given, simply introduce equations of the form $x_k+b_k=0\text{ mod 2}$ for each $k=0,...,\ell-1$. For equality constraints we have $\xi_\ell^==\ell$ equations and $\eta_\ell^==\ell$ equations that can be maximally satisfied.

\subsection{Embedding a complete ILP instance into max-XORSAT} \label{subsec:obj}

The basic idea here is analogous to that of binary search in the literature of MaxSAT solvers \cite{Ansotegui2013SATBasedmaxSAT,Morgado2013IterativeCoreGuided}, where a SAT oracle is used to iteratively find the optimum. More specifically, to solve the problem in \eqref{eq:01ILP}, consider a formulation of constraints (call it ILP-c) with lower bound $\beta$ on the objective function
\begin{equation}\label{eq:ilp_beta}
\begin{aligned}
{\bf c}^T{\bf x} &\ge \beta \\
A{\bf x} &< {\bf b}+{\bf 1} \\
A_={\bf x} &= {\bf b}_=
\end{aligned}
\end{equation}
for ${\bf x}\in\{0,1\}^n$ and $\beta$ being an integer less than $\|{\bf c}\|_1$. Then the optimization problem in \eqref{eq:01ILP} can be solved by finding the critical value $\beta^*$ such that \eqref{eq:ilp_beta} is feasible when $\beta=\beta^*$ but not feasible when $\beta=\beta^*+1$. The feasibility of \eqref{eq:ilp_beta} can be determined by checking the total number of equations in the corresponding max-XORSAT instance. Recall that $A\in\mathbb{R}^{m\times n}$, ${\bf b}\in\mathbb{R}^m$, $A_=\in\mathbb{R}^{p\times n}$, and ${\bf b}_=\in\mathbb{R}^p$. Suppose each element $c_i$ of ${\bf c}$ uses $\gamma_i$ bits, and the element $(i,j)$ of $A$ and $A_=$ requires $\alpha_{ij}$ and $\alpha_{ij}^=$ bits respectively. By application of the results in Section \ref{subsec:constraints} in embedding pseudo-Boolean constraints, we have that the total number of equations in ILP-c is

\begin{equation}\label{eq:xi_ilp_c}
\begin{array}{ccl}
    \xi^\text{ILP-c} & = & \displaystyle 
    \underbrace{
    \xi_{\gamma_1,\gamma_2,\cdots,\gamma_n}^\text{MIA}+\xi_{\mu_{J(n)+1,1}
    (\gamma_1,\gamma_2,\cdots,\gamma_n)}^\ge 
    }_{{\bf c}^T{\bf x}\ge \beta}
    + \underbrace{\sum_{i=1}^m\left(\xi_{\alpha_{i1},\alpha_{i2},\cdots,\alpha_{in}}^\text{MIA}+\xi_{\mu_{J(n)+1,1}(\alpha_{i1},\alpha_{i2},\cdots,\alpha_{in})}^<\right)}_{A{\bf x}<{\bf b}+{\bf 1}} \\
    & + & \displaystyle \underbrace{\sum_{i=1}^p\left(\xi_{\alpha_{i1}^=,\alpha_{i2}^=,\cdots,\alpha_{in}^=}^\text{MIA}+\xi_{\mu_{J(n)+1,1}(\alpha_{i1}^=,\alpha_{i2}^=,\cdots,\alpha_{in}^=)}^=\right)}_{A_={\bf x}={\bf b}_=}.
\end{array}
\end{equation}
where contributions from each component of ILP-c in \eqref{eq:ilp_beta} is indicated with curly brackets. Similarly, the maximum number of equations that can be satisfied $\eta^\text{ILP-c}$ can be calculated by substituting $\xi$ with $\eta$ in \eqref{eq:xi_ilp_c}.

It would be useful to estimate the total number of equations and variables contained in the max-XORSAT instance generated from the ILP-c problem \eqref{eq:ilp_beta}. To simplify the analysis, define $\alpha=\max_{i,j}\{\alpha_{ij}\}$, $\alpha^==\max_{i,j}\{\alpha^=_{ij}\}$, and $\gamma=\max_i\{\gamma_i\}$. Here we show that
\begin{equation}
    \xi^\text{ILP-c}\in O(n(\gamma+m\alpha+p\alpha^=))\subseteq\tilde{O}(mn).
\end{equation}

From \eqref{eq:xi_wia} we have $\xi_\ell^\text{WIA}\le1+14\ell$. Similarly, from \eqref{eq:xi_ia} we have $\xi_\ell^\text{IA}\le 14\ell$. Applying these upper bounds, combined with $\xi_\ell^\text{HWA}=\ell$ from previous discussions, leads to an upper bound for $\xi_{\alpha_1,\alpha_2,\cdots,\alpha_n}^\text{MIA}$ as the follows
\begin{equation}\label{eq:xi_ub}
\begin{array}{ccl}
    \xi_{\alpha_1,\alpha_2,\cdots,\alpha_n}^\text{MIA} & \le & \displaystyle \sum_{k=1}^{\lceil n/2\rceil}(1+14\mu_{1,k}(\alpha_1,\alpha_2,\cdots,\alpha_n)) + \alpha + \sum_{j=2}^{J(n)+1}\sum_{k=1}^{m_j}14\mu_{j,k}(\alpha_1,\alpha_2,\cdots,\alpha_n) \\
    & = & \displaystyle \alpha+\lceil\frac{n}{2}\rceil + 14
    \underbrace{\sum_{j=1}^{J(n)+1}\sum_{k=1}^{m_j}\mu_{j,k}(\alpha_1,\alpha_2,\cdots,\alpha_n)}_{(*)}.
\end{array}
\end{equation}
Here the factor 14 that appears in the upper bound comes from the 14 equations that the CARRY gadget requires. The sum $(*)$ in \eqref{eq:xi_ub} over $\mu_{j,k}$ is eseentially over the nodes $w_{j,k}$ in the binary tree of integer adders (IA) contained in the multiple integer adder (MIA) construction (Figure \ref{fig:mia-example} shows an example), which can be bounded from above by considering the case of a full binary tree of depth $\delta=\lceil\log_2n\rceil$. Observing that $\mu_{j,k}\le\alpha_j$ for any $j=1,\cdots,\delta$, we have that the sum $(*)$ is bounded from above by
\begin{equation}
\sum_{j=1}^\delta(\alpha+j)m_j=\alpha(2^{\delta+1}-2)+2^{\delta+2}-2(\delta+2)\le (2\alpha+4)2^{\lceil\log_2 n\rceil}\in O(\alpha n).
\end{equation}
The same upper bound argument can be also applied to the integer comparators to yield
\begin{equation}
    \xi_{\mu_{J(n)+1,1}(\alpha_1,\alpha_2,\cdots,\alpha_n)}^<\le\xi_{\mu_{\delta,1}(\alpha_1,\alpha_2,\cdots,\alpha_n)}^<\le 5\mu_{\delta,1}(\alpha_1,\alpha_2,\cdots,\alpha_n)=5(\alpha+\delta).
\end{equation}
Similarly we have $\xi_{\mu_{J(n)+1,1}(\alpha_1,\alpha_2,\cdots,\alpha_n)}^\ge \le 5(\alpha+\delta)$. The upper bounds of the number of equations in MIAs and integer comparators now allow us to find an upper bound to $\xi^\text{ILP-c}$ as
\begin{equation}\label{eq:xi_ilp_c_ub}
    \begin{array}{ccl}
       \xi^\text{ILP-c} & \le & 14(2\gamma+4)2^{\lceil\log_2n\rceil} + \lfloor n/2\rfloor + \gamma + 5(\gamma+\lceil\log_2 n\rceil) \\
         & + & m\cdot(14(2\alpha+4)2^{\lceil\log_2n\rceil} + \lfloor n/2\rfloor + \alpha + 5(\alpha+\lceil\log_2 n\rceil)) \\
         & + & p\cdot(14(2\alpha^=+4)2^{\lceil\log_2n\rceil} + \lfloor n/2\rfloor + \alpha^= + 5(\alpha^=+\lceil\log_2 n\rceil)) \\
         & & \in O(n(\gamma+m\alpha+p\alpha^=))\subseteq\tilde{O}(mn).
    \end{array}
\end{equation}
The same scaling can be found with the maximum number of equations that can be satisfied, $\eta^\text{ILP-c}$. A stronger observation can be made from \eqref{eq:xi_ilp_c_ub} that the dominant contribution in the upper bound taking the form of $14(2\gamma+2m\alpha+2p\alpha^=)2^{\lceil\log_2 n\rceil}$ comes exclusively from the CARRY gadgets. This observation also allows the conclusion that the total number of variables in the max-XORSAT instance, which will also be dominated by the carry bits and auxiliary bits from the CARRY gadgets, is also $O(n(\gamma+m\alpha+p\alpha^=))\subseteq\tilde{O}(mn)$.

\subsection{From max-XORSAT to LDPC code}

Let $B{\bf y}={\bf s}$ be the max-XORSAT instance that comes from ILP-c described in Section \ref{subsec:obj}, and $B$ is a matrix of size $\xi^\text{ILP-c}\times n'$ with $n'\in\tilde{O}(mn)$ being the total number of bits in the max-XORSAT instance. For characterizing the extent of possible quantum advantage in DQI, it is important to understand the quality of the LDPC code $\mathcal{C}^\perp=\{{\bf x'}\in\{0,1\}^{n'}|B^T{\bf x'}=0\}$ coming from the max-XORSAT instance. A key parameter that determines the prospect of algorithmic advantage in DQI is the code distance $d$ of $\mathcal{C}^\perp$, which is equivalent to the minimum number of rows in $B$ that are linearly dependent.

With the construction described in Section \ref{subsec:constraints}, one can see from the construction of the CARRY gadget (Equation \ref{eq:carry_gadget}) that the following equations sum to 0 mod 2:
\begin{equation}\label{eq:CARRY_old}
    \begin{array}{r}
         v_k+u_k=0 \\
         v_k+c_{k-1}=0 \\
         u_k+c_{k-1}=0.
    \end{array}
\end{equation}
This, combined with the fact that no subset of two equations in any of the max-XORSAT construction are linearly dependent, solidifies that the code distance $d=3$ regardless of the dimensions of $B$. Such a code distance barely qualifies $\mathcal{C}^\perp$ as an error correcting code, and there should be ways to improve it. One strategy is to consider an alternative max-XORSAT encoding of the carry logic $c_k=\text{maj}(u_k,v_k,c_{k-1})=u_k\cdot v_k+u_k\cdot c_{k-1}+v_k\cdot c_{k-1}$ as the following:
\begin{equation}\label{eq:CARRY_alt}
    \begin{array}{r}
         c_k+u_k=0 \\
         c_k+v_k=0 \\
         c_k+c_{k-1}=0.
    \end{array}
\end{equation}
Let $F(c;u_k,v_k,c_{k-1})$ be the number of equations satisfied given an assignment of $u_k$, $v_k$ and $c_{k-1}$ values and $c_k=c$. Then it can be checked that $c_k=\text{argmax}_{c\in\{0,1\}}F(c;u_k,v_k,c_{k-1})=\text{maj}(u_k,v_k,c_{k-1})$. The alternative version of the CARRY logic in \eqref{eq:CARRY_alt} avoids having 3 linearly dependent rows in $B$ and potentially improves the code distance to $d\ge 4$. However, for this study we still adopt the construction in \eqref{eq:CARRY_old} for numerical experiments. The same experiments for improved versions of the LDPC code is left for future work.

\section{Methods}\label{sec:methods}
\subsection{Classical solvers}
\label{sec:classical_solvers}
As one of the world's most powerful classical optimization solvers, Gurobi~\cite{gurobi} produces optimal solutions for all considered problem instances in this work, and is used as baseline to compare the performances of the quantum solvers against. It outputs sets of solution variables $\{x_i\}$ and satisfied constraints $S_i$ that solve each of the considered problems optimally. We formulate the underlying optimization problem in Gurobi in two different ways: As an ILP, without transformations, and as a max-XORSAT problem by formulating the problem as a SAT problem instance with additional constraints. The details of both formulations are described in the following.

\subsubsection{Problem Formulation in Gurobi as an Integer Linear Program}
\label{sec:ilp_Gurobi}

In 0-1 ILP formulation, the problem is represented by a linear objective function that aims to maximize or minimize a specific outcome, as described in Eq.~\eqref{eq:ilp_objective}, subject to a set of linear constraints, Eq.~\eqref{eq:ilp_constraints}. Each decision variable is defined as a binary variable, taking values of either 0 or 1. The objective function is constructed by assigning coefficients to these variables, which represent their contribution to the overall goal. The constraints are formulated as linear inequalities, ensuring that the solution adheres to the specified limits and conditions of the problem. This straightforward representation allows for efficient optimization using Gurobi's powerful solver, which employs advanced algorithms to explore feasible solutions.

\subsubsection{Problem Formulation in Gurobi as a max-XORSAT Problem}
\label{sec:max-xor-sat_Gurobi}
In the max-XORSAT formulation, the problem is specifically tailored to handle XOR constraints, which are common in logical optimization scenarios. The encoding begins by defining binary variables that represent the components of each XOR constraint. Each constraint $i\in m$ is structured such that the sum of the involved decision variables $\textbf{b}_i\textbf{x}$ must satisfy specific parity conditions (see Eq.~\eqref{eq:max_xorsat_objective}: it must be odd if the associated constraint of the satisfiability problem $\textbf{v}_i=1$ and even if $\textbf{v}_i=0$). This logical representation allows for a more direct and efficient modeling of problems where XOR relationships are critical, enabling Gurobi to optimize the number of satisfied constraints effectively. The formulation also introduces auxiliary variables to facilitate the enforcement of these parity conditions. By incorporating these auxiliary variables, the model can maintain the integrity of the XOR relationships while still adhering to the requirements of a mixed-integer programming framework. The objective function is then defined to maximize the total number of satisfied XOR constraints, which represents a max-XORSAT problem.

\subsection{Soft Belief Propagation (BP2)}
We saw in section \ref{sec:gallager_bf} that the hard-decision binary belief propagation algorithm, BP1, has a somewhat limited effectivity for the type of max-XORSAT problems we consider in this work. In this section, we investigate a state-of-the-art belief propagation, BP2, algorithm which does not rely on a hard-decision threshold in order to correct the errors in codeword $\mathbf{y}$, see for example \cite{10.5555/1795974} and references therein. The algorithm uses as input the log-likelihood ratios (LLRs) of the noisy code word $\mathbf{y}$, which we assume to have gone through a binary symmetric channel (BSC) with error rate $p=0.001$, see for example \cite{748992}, where also Gaussian channels are shown. Empirically, this choice seems to perform best for our application. 

\begin{figure}[h]
    \centering
    \includegraphics[width=0.9\linewidth]{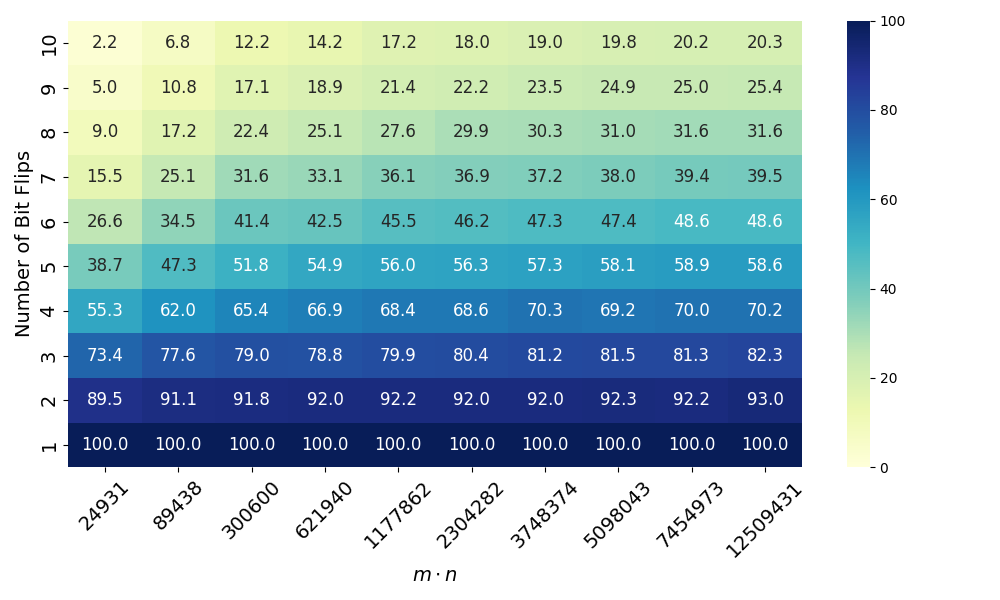}
    \caption{Success rate of our implementation of a state-of-the-art, soft-decision, belief propagation decoder (BP2) for the zero codeword with $\ell$ randomly generated  bit flips as a function of $\ell$ and the problem size, i.e, the size of $B$. We set $T=5$. We use $10^4$ sampled code words for each value of $\ell$ and $B$. }
    \label{fig:benchmark_sota_bp}
\end{figure}

In Algorithm \ref{alg:sota_bp}, we show the pseudo code of the  belief propagation we use in our comparisons. Note that the algorithm does not make the decision of which bits to flip based on a hard threshold, instead, it softens the decision by using the provided likelihoods of the assumed channel. This has the effect to greatly improve its performance as seen in Fig. \ref{fig:benchmark_sota_bp} as compared to Fig. \ref{fig:benchmark_gallager_bp}. As before, and using the exact same parity-check matrices $B^T$ as in Fig.~\ref{fig:benchmark_gallager_bp}, we show an estimate of the success rate for decoding the zero codeword $\mathbf{c}$ with $\ell$ random bit flips as a function of both, the number of errors and the problem size. We use this belief propagation algorithm to compare what is possible with DQI in principle, if a quantum-circuit version of this algorithm were to be implemented. While such a quantum implementation does not yet exist, Refs.~\cite{Renes_2017, Piveteau_2022} suggest promising directions that could be pursued to develop one. However, constructing this quantum version is beyond the scope of the present work.

\begin{algorithm}
\caption{Belief Propagation for LDPC Decoding}
\begin{algorithmic}[1]
\STATE \textbf{Input:} Parity-check matrix $B^T \in \{0,1\}^{m \times n}$, received bits $y \in \{0,1\}^n$, max iterations $T$
\STATE \textbf{Output:} Decoded bit vector $\texttt{decoded\_bits}$ or decoding failure

\STATE Set channel crossover probability $p \gets 0.001$
\STATE $\texttt{received\_llrs} \gets \log\left(\frac{1-p}{p}\right) \cdot (1 - 2y)$
\STATE Initialize message matrix $M^{(v \to c)} \gets$ replicate $\texttt{received\_llrs}$ across $m$ rows
\STATE $n \gets$ number of variable nodes (columns), $m \gets$ number of check nodes (rows)

\FOR{$t = 0$ to $T-1$}
    \STATE Initialize $M^{(c to v)} \gets 0$ matrix of same shape as $M^{(v \to c)}$
    \FOR{each check node $i = 0$ to $m-1$}
        \STATE $\texttt{connected\_vnodes} \gets \{j \mid B^T_{ij} = 1\}$
        \FOR{each variable node $v \in \texttt{connected\_vnodes}$}
            \STATE $P \gets \prod_{j \in \texttt{connected\_vnodes} \setminus \{v\}} \tanh\left(0.5 \cdot M^{(v \to c)}[i, j]\right)$
            \STATE $M^{(c \to v)}[i, v] \gets 2 \cdot \tanh^{-1}(P)$
        \ENDFOR
    \ENDFOR

    \FOR{each variable node $j = 0$ to $n-1$}
        \STATE $\texttt{connected\_cnodes} \gets \{i \mid B^T_{ij} = 1\}$
        \FOR{each check node $c \in \texttt{connected\_cnodes}$}
            \STATE $M^{(v \to c)}[c, j] \gets \texttt{received\_llrs}[j] + \sum_{i \in \texttt{connected\_cnodes} \setminus \{c\}} M^{(c \to v)}[i, j]$
        \ENDFOR
    \ENDFOR

    \STATE $\texttt{final\_llrs} \gets \texttt{received\_llrs} + \sum_{i=0}^{m-1} M^{(c \to v)}[i, :]$
    \STATE $\texttt{decoded\_bits} \gets 1\; \{\texttt{final\_llrs} < 0\}$ \COMMENT{1 if negative LLR, else 0}

    \IF{$B \cdot \texttt{decoded\_bits} \bmod 2 = 0$}
        \STATE \textbf{return} $\texttt{decoded\_bits}$ \COMMENT{Successful decoding}
    \ENDIF
\ENDFOR

\STATE \textbf{return} failure with $\texttt{decoded\_bits}$ \COMMENT{max iterations reached}
\end{algorithmic}\label{alg:sota_bp}
\end{algorithm}

\subsection{Performance metrics}
\label{sec: performance metrics}
In this section, we closely follow Ref.~\cite{jordan2024optimizationdecodedquantuminterferometry} to describe the behavior of the DQI algorithm in the case of imperfect decoding and then we comment on how to interpret the results we show in the next section. In practice, the quantum state after the decoding state will not be the one in Eq.~\eqref{eq:decoded_state} due to imperfect decoding. Instead, one obtains

\begin{equation}
    \sum_{k=0}^\ell\omega_k\frac{1}{\sqrt{\binom{m}{k}}}\left(\sum_{\mathbf{y}\in\mathcal{D}_k} (-1)^{\mathbf{v}\cdot\mathbf{y}} |0^{\otimes m}\rangle|B^T\mathbf{y}\rangle+ \sum_{\mathbf{y}\in\mathcal{F}_k} (-1)^{\mathbf{v}\cdot\mathbf{y}} |\mathbf{y}\oplus\mathbf{y'}\rangle|B^T\mathbf{y}\rangle\right)
\end{equation}
where $\mathbf{y}\neq\mathbf{y'}$ and  $\mathcal{D}_k$ ($\mathcal{F}_k$) is the set of $\mathbf{y}\in\{0,1\}^m$ such that $|\mathbf{y}|=k$ for which the belief propagation algorithm succeeded (failed). Note that if the belief propagation algorithm fails to decode a word with error rate $\epsilon_k$, then $|\mathcal{F}_k|=\epsilon_k\binom{m}{k}$. 

In the DQI protocol, one measures the \emph{message} register and post-selects the results for which the all-zero state is measured. The renormalized resulting state
\begin{equation}
    \frac{1}{R}\sum_{k=0}^\ell\omega_k\frac{1}{\sqrt{\binom{m}{k}}}\sum_{\mathbf{y}\in\mathcal{D}_k} (-1)^{\mathbf{v}\cdot\mathbf{y}} |B^T\mathbf{y}\rangle,
\end{equation}
where $R = \sum_{k=0}^\ell\omega_k^2(1-\epsilon_k)$, replaces the ideal state in Eq.~\eqref{eq:decoded_state} and leads to a modified DQI state, $|\widetilde{DQI}\rangle$, after application of the Hadamard gates as indicated in Eq.~\eqref{eq:dqi_state}. Note that $R$ denotes the probability of measuring the \emph{message} register in the all-zero state. Consequently, the number of post-selected shots is equal to the total number of shots multiplied by $R$. Therefore, the decoder's success rate $(1 - \epsilon_k)$ (benchmarked in Figs.~\ref{fig:benchmark_gallager_bp} and~\ref{fig:benchmark_sota_bp}) directly influences the number of post-selected shots and, in turn, determines the overall sampling complexity of DQI.

When quantum hardware is available at scale, it will be possible to directly measure the imperfect DQI state, $|\widetilde{DQI}\rangle$, and obtain a solution of the optimization problem. In the meantime, for simulations and for asymptotic behavior, it is only possible to estimate its expected value explicitly, which does not constitute a solution of the optimization problem. The observable
\begin{equation}
    O = \sum_{i=1}^m (-1)^{v_i}\prod_{j:B_{ij}=1}^nZ_j
\end{equation}
represents the max-XORSAT objective function Eq.~\eqref{eq:max_xorsat_objective}. We can compute the expected value of this observable with respected to the imperfect DQI state and obtain
\begin{equation}
    \langle O\rangle = \langle \widetilde{DQI}|O|\widetilde{DQI}\rangle = \sum_{k,k'}^\ell\omega_k\omega_{k'}A_{kk'}^{(m,\ell)}
\end{equation}
where
\begin{equation}
\label{eq: matrix A}
    A_{kk'}^{(m,\ell)} = \frac{1}{R^2}\frac{1}{\sqrt{\binom{m}{k}\binom{m}{k'}}}\sum_{\mathbf{y}\in\mathcal{D}_k}\sum_{\mathbf{y'}\in\mathcal{D}_{k'}}(-1)^{\mathbf{v}\cdot(\mathbf{y}+\mathbf{y'})}\sum_{i=1}^m (-1)^{v_i}\langle B^T\mathbf{y}|B^T(\mathbf{y'}+\mathbf{e}_i)\rangle
\end{equation}
where $\mathbf{e}_i\in\{0,1\}^m$ are vectors such that their entries $e_{i}[k]$ are equal to $1$ when $i=k$ and $0$ otherwise.

From this expected value, it is easy to construct the quantity of interest in our case: the number of fulfilled constraints $S$, which we can only know in expectation
\begin{equation}\label{eq:expected_sat_constraints}
    \langle S \rangle = \frac{1}{2}\left(\langle O \rangle + m\right).
\end{equation}
It is clear that these equations scale unfavorably with the number of constraints $m$, which limits their scope, i.e., the size of problems we can investigate, as we will see in our experiments. We will use Eq.~\eqref{eq:expected_sat_constraints} as our main performance metric and will compare it to the classical solution of our optimization problems obtained using Gurobi.

\subsection{Logical quantum resource estimation}
\label{sec: resource estimation}
To estimate the quantum resources required for implementing the DQI algorithm at the level of logical qubits and gates, we evaluate both the total number of qubits and the number of each gate type used. The total number of qubits can be determined precisely by tracking the ancillary registers introduced in the BP1 decoding quantum circuit in Sec.~\ref{sec:quantum_bf_gallager}. This is because the decoding subroutine is the only component of the DQI algorithm that requires the addition of auxiliary qubits. Specifically, up to $m$ qubits are needed to prepare the Dicke states, which corresponds to the number of qubits in the \emph{message} register, i.e, the Dicke state preparation does not increase the qubit count of the total circuit. 

With these considerations, we find that the total number of qubits, $N_q$, is given by:
\begin{equation}
\label{eq: number of qubits}
N_q = m + n + 2\left\lceil \log_2(t + 1) \right\rceil + T \cdot m + (T - 1) \cdot n 
= (T + 1) \cdot m + T \cdot n + 2\left\lceil \log_2(t + 1) \right\rceil,
\end{equation}
where:
\begin{itemize}
    \item $m + n$ accounts for the \emph{message} and \emph{syndrome} registers.
    \item $2\left\lceil \log_2(t + 1) \right\rceil$ covers the \emph{hamming} and \emph{comparator} registers, with $t$ being the maximum number of variables in any constraint.
    \item $T \cdot m$ represents the \emph{flip} registers created during each iteration,.
    \item $(T - 1) \cdot n$ includes additional \emph{syndrome} registers appended at the end of each iteration except the for the last one. 
\end{itemize}

Since $t \leq n$, the total number of qubits grows linearly with both the number of constraints and number of variables, and  also with the number of iterations. Also, the number of qubits does not depend on the number of errors we ask the decoder to correct, $\ell$.  In the results section, we validate the correctness of this expression by directly counting the number of qubits in the quantum circuit constructed for each specific instance.

To evaluate the number of gates used in the full DQI circuit, we first select a basis gate set into which we transpile the circuit. Following Ref.~\cite{patamawisut2025quantumcircuitdesigndecoded}, we choose the gate set $\{\text{Z}, \text{CNOT}, \text{RX}, \text{RY}, \text{RZ}, \text{SWAP}\}$. The transpilation is performed using the \texttt{transpile} function from Qiskit~\cite{javadiabhari2024quantumcomputingqiskit}. However, due to the high depth of circuits resulting from the max-XORSAT instances we consider, transpiling the entire DQI circuit in one step becomes computationally infeasible even for moderately sized problems. 
Instead, we transpile each modular block of the circuit individually and aggregate their gate counts. This approach yields an upper bound on the total number of gates, as transpiling the entire circuit at once would allow for further gate cancellations and simplifications across blocks. Nevertheless, for the purpose of this work—namely, estimating resource usage and verifying that the gate count scales polynomially—we find that block-wise transpilation provides a sufficient  estimation. More concretely, the circuit blocks that are individually transpiled, as described in Sec.~\ref{sec:preliminary_quantum_circuit_blocks}, include:
\begin{itemize}
    \item The unary encoding step.
    \item The Dicke state preparation from unary-encoded states.
    \item The Encoding phases block, which is trivially transpiled since it consists of non-canceling Z gates acting on different qubits.
    \item The syndrome encoding step, which involves non-overlapping CNOT gates (i.e., no two CNOTs share the same control and target qubits within the same step) and is also trivially transpiled. 
\end{itemize}
For the BP1 quantum circuit, Sec.~\ref{sec:quantum_bf_gallager}, we also perform transpilation block-wise. Specifically:
\begin{itemize}
    \item The controlled $U_{+1}$ gate used in Algorithm~\ref{alg:increment_by_one} is inserted as a pre-transpiled module.
    \item The \texttt{IntegerComparator} gate is similarly inserted in its already transpiled form.
    \item Both the inverses of the controlled $U_{+1}$ and the \texttt{IntegerComparator} gates are also used in their transpiled versions.

\end{itemize}
Finally, the Hadamard layer applied at the end of the DQI circuit is transpiled as a separate block as well. This modular transpilation strategy ensures manageable compilation while still providing a meaningful upper-bound estimate for the total gate complexity of the DQI circuit. After constructing the circuit using the block-wise transpilation approach described above, the number of gates of each type can be directly extracted from the resulting quantum circuit, which contains only gates from the chosen basis set. In the results section, we show the number an types of gate used for different problem sizes and verify a polynomial scaling with $m$, $n$, $\ell$ and $T$. 

\subsection{Experiment design}
\label{sec: experiment design}

For our numerical studies, we want to achieve three objectives: 1. show how DQI performs on average for problems of increasing complexity, 2. show that our quantum circuit implementation matches the analytical performance of DQI, which can only be demonstrated for small systems, and 3. estimate the quantum resources needed to run it. We detail below the types of experiments we perform:
\begin{itemize}

    \item \textbf{Average performance scaling:} We generate an instance of the ILP problem corresponding to the automotive bundling task described in Sec.~\ref{sec:problem_formulation}. We transform the smallest relevant instance of this problem into a max-XORSAT instance, following the procedure outlined in Sec.~\ref{sec: trasnformation to max-XOR-Sat}. The resulting max-XORSAT problem contains $m = 827$ constraints and $n = 345$ variables, yielding a total problem size of $m \cdot n = 285315$. This instance already produces a quantum circuit that is too large for classical simulation. Moreover, estimating $\langle S \rangle$ classically via Eq.~\eqref{eq: matrix A} also becomes computationally infeasible. To overcome these limitations and to benchmark problem instances that still resemble the industrially relevant ILP, we generate smaller instances of the matrix $B$ using the sampling procedure described in Appendix~\ref{app:sampling_B}. This allows us to create problem sizes ranging from $m \cdot n = 276$ up to $m \cdot n = 10296$.

    \item \textbf{Quantum implementation test:} We simulate the full quantum circuit execution of DQI for random sparse max-XORSAT problems that result in quantum circuits with fewer than 30 qubits. Specifically, we generate random instances of $B$ and $\mathbf{v}$ and simulate the execution of the proposed quantum circuit for DQI with the following parameters: 
\begin{itemize}
  \item $m=8$, $n=6$, $\ell=2$, $T=1$ (50 random instances);
  \item $m=6$, $n=4$, $\ell=1$, $T=1$ (50 random instances);
  \item $m=5$, $n=3$, $\ell=1$, $T=3$ (50 random instances).
\end{itemize}

This allows us to verify that the predicted number of satisfied constraints matches the average obtained from direct sampling of the quantum circuit.
 This validation builds confidence in our quantum circuit implementation and the reliability of predicted results for larger, unsimulable instances. Moreover, in these smaller cases, we are not limited to computing only the expected number of satisfied constraints; we can also extract the full histogram of solutions and identify optimal solutions sampled by the DQI algorithm. 
\end{itemize}

For all the experiments listed above, we evaluate the expected number of satisfied constraints as a primary performance metric and estimate the quantum resources required by our quantum circuit implementation. Importantly, thanks to the block-wise transpilation strategy described in Sec.~\ref{sec: resource estimation}, we can estimate the quantum resources required for arbitrarily large problem instances. In addition, we also assess both performance and resource requirements using different values of $\ell$, the maximum number of errors we require the decoder to correct, as well as varying the number of iterations of the belief propagation algorithms we test.

For all instances where the performance is evaluated using the number of satisfied constraints, we compare the results against the optimal solution obtained by transforming the max-XORSAT problem into a linear optimization problem which we solve with Gurobi. This transformation is carried out using the methods described in Sec.~\ref{sec:classical_solvers}. The optimal solution returned by Gurobi serves as reference for the maximum number of satisfiable constraints. Additionally, we benchmark our results against a baseline obtained by randomly sampling bitstrings, providing a reference for evaluating the improvement of the DQI algorithm when compared to random sampling.

\section{Results and Analysis}
\label{sec: results}
\subsection{Comparative Solver Performance}

\begin{figure}[H]
    \centering
    \includegraphics[width=0.9\linewidth]{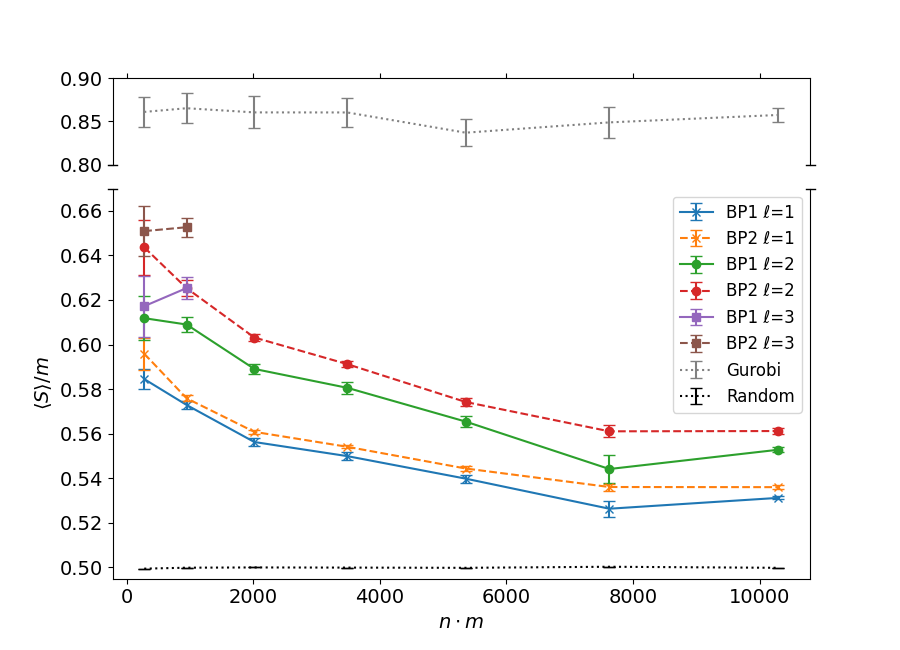}
    \caption{Performance comparison of DQI for downscaled max-XORSAT instances derived from the automotive bundling ILP. Each data point corresponds to the average fraction of satisfied constraints $\langle S\rangle/m$ over 5 independently sampled problem instances. We evaluate DQI using both BP1 and BP2 decoders with $T=5$ iterations, and up to $\ell=3$. The results are compared against random sampling and the optimal solution computed using Gurobi. 
 }
    \label{fig: performance reduced  ILP}
\end{figure}

We evaluate the performance of DQI using both algorithms BP1 and BP2 with $T=5$. The DQI performance is tested up to $\ell=3$, as the computational cost of obtaining $\langle S\rangle$ becomes infeasible for larger $\ell$. In Fig.~\ref{fig: performance reduced  ILP}, we show the performance of our DQI implementation for different sized sampled instances of a max-XORSAT problem derived from the automotive bundling ILP described in Sec.~\ref{sec:problem_formulation}.

We generate 5 different sets of down-scaled max-XORSAT problems, according to the procedure described in Appendix~\ref{app:sampling_B}, and repeat the entire experiment for each set. The fraction of satisfied constraints is plotted as the average over the 5 sets, for DQI, random sampling, and the Gurobi results. We observe that the average performance of the DQI algorithm decreases as the problem size increases, eventually appearing to converge toward an asymptotic value. Across all tested instances, DQI consistently outperforms random sampling. However, compared to the state-of-the-art classical solver \texttt{Gurobi}, the performance gap remains significant. Note that the Gurobi results are the optimal solutions of the problem, which we are comparing to the average over all possible bit strings sampled by DQI. This fact could be responsible for the observed gap.

We also see that, for the same value of $\ell$, the BP1 decoder performs worse than BP2. This observation is consistent with the success rates shown in Fig.~\ref{fig:benchmark_gallager_bp}, where the BP2 decoder clearly achieves higher success rates. Furthermore, increasing $\ell$ in the DQI algorithm improves the number of satisfied constraints. We expect this trend to continue for larger values of $\ell$ not tested in our study, provided the decoder’s success rate continues reliably decoding larger numbers of errors $\ell$.

\subsection{Resource Estimation }
We present the quantum resources—namely, the number of qubits and the number of gates—required for the DQI circuit implementation using the BP1 decoder. These resources are evaluated for the same down-scaled max-XORSAT instances, as those used in our performance study. Unlike in the performance evaluation, the resource estimation is not limited by computational constraints associated with problem size. This allows us to clearly illustrate the observed scaling behavior. Consequently, we evaluate the quantum resources for problems up to $m \cdot n = 116886$ instead of the $m \cdot n = 10296$ limit reached in the performance estimation.

\begin{figure}[H]
    \centering
    \includegraphics[width=0.9\linewidth]{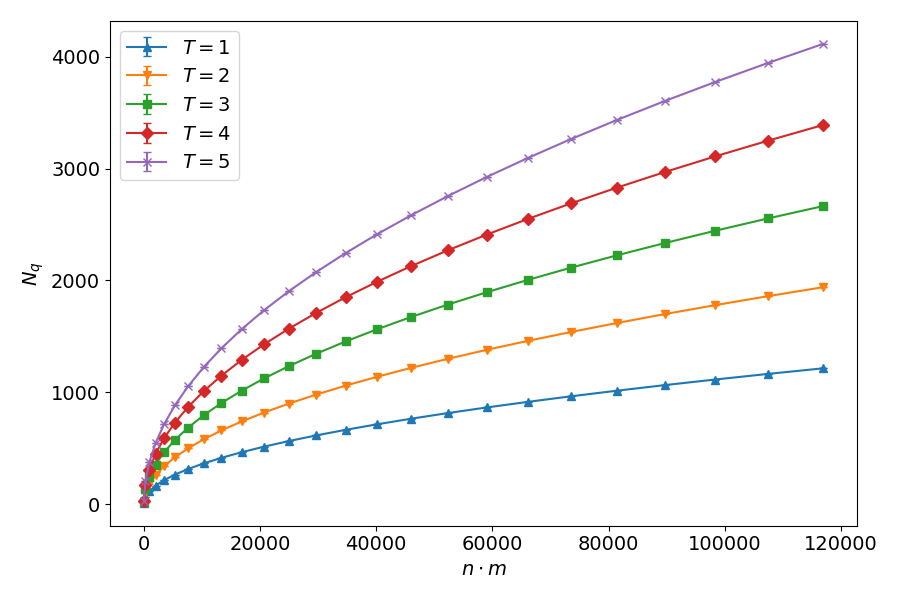}
    \caption{Number of qubits required to implement the  proposed DQI circuit for downscaled max-XORSAT instances derived from the automotive bundling ILP. Each data point corresponds to the average number of qubits over 5 independently sampled problem instances. The error bars are included in the plot, but they are not visible due to their small magnitude relative to the scale of the y-axis.  } 
    \label{fig:qubit_count}
\end{figure}

 As in the performance estimation, we generate 5 different sets of down-scaled max-XORSAT problems and repeat the entire experiment for each set.  The plotted results represent the average over these five instances, and the corresponding error bars indicate the standard error of the mean. In Fig.~\ref{fig:qubit_count}, we show the number of qubits required by the DQI circuit as a function of the problem size. This quantity can be obtained either by directly counting the qubits in the quantum circuit object or by evaluating Eq.~\eqref{eq: number of qubits}. As predicted by Eq.~\eqref{eq: number of qubits}, the number of qubits scales linearly with both $m$ and $n$. This results in a square root scaling with $m \cdot n$, as observed in the plot. Moreover, the number of qubits is independent of the maximum number of error, $\ell$, required to decode and increases linearly with the number of iterations $T$. Since $m$ and $n$ are fixed across all five instances tested for each problem size, the variance arises solely from variations in the maximum number of variables involved in any constraint, ($t$ in Eq.~\eqref{eq: number of qubits}), which may differ among the 5 sets of max-XORSAT problems tested.

\begin{figure}[H]
    \centering
    \includegraphics[width=0.9\linewidth]{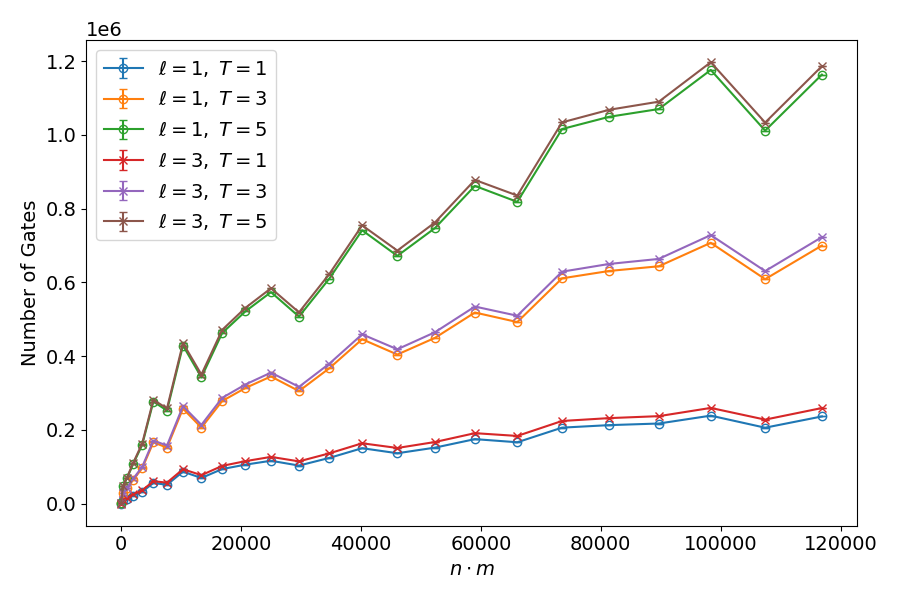}
    \caption{Total number of gates required to implement the  proposed DQI circuit after transpilation by blocks into the gate set $\{\text{Z}, \text{CNOT}, \text{RX}, \text{RY}, \text{RZ}, \text{SWAP}\}$.Each data point corresponds to the average number of gates over 5 independently sampled problem instances. The error bars are included in the plot, but they are not visible due to their small magnitude relative to the scale of the y-axis.  }
    \label{fig:gate_count}
\end{figure}

\begin{figure}[h]
    \centering\includegraphics[width=0.9\linewidth]{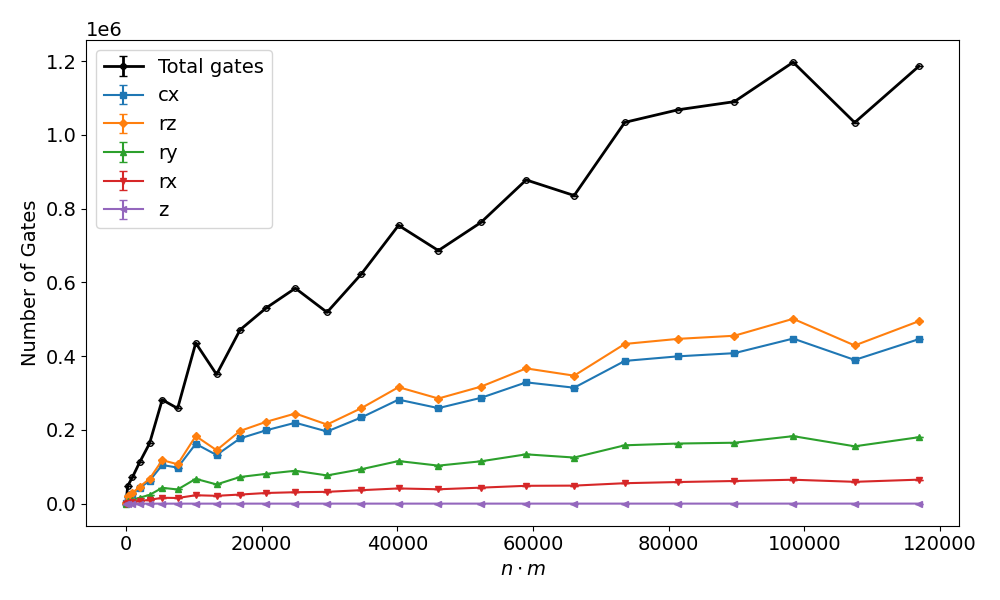}
    \caption{Total number of gates of each type required to implement the  proposed DQI circuit after transpilation by blocks into the gate set $\{\text{Z}, \text{CNOT}, \text{RX}, \text{RY}, \text{RZ}, \text{SWAP}\}$. Each data point corresponds to the average over 5 independently sampled problem instances. We set $\ell=3$ and $T=5$. The error bars are included in the plot, but they are not visible due to their small magnitude relative to the scale of the y-axis.    }
    \label{fig: gate_breakdown}
\end{figure}

In Fig.~\ref{fig:gate_count}, we show the total number of gates in the DQI circuit as a function of problem size. This gate count is obtained by transpiling the circuit block-by-block into the gate set $\{\text{Z}, \text{CNOT}, \text{RX}, \text{RY}, \text{RZ}, \text{SWAP}\}$, as described in Sec.~\ref{sec: resource estimation}. Additionally, in Fig.~\ref{fig: gate_breakdown} we show the individual scaling of each gate among the gate set $\{\text{Z}, \text{CNOT}, \text{RX}, \text{RY}, \text{RZ}, \text{SWAP}\}$ for $\ell=3$ and $T=5$. As before, we repeat the experiment for the 5 sets of max-XORSAT problems tested before and plot the average gate count along with the associated statistical error bars (standard error of the mean). We observe that the total number of gates exhibits a similar scaling trend to that of the number of qubits with respect to the problem size $m \cdot n$, indicating sublinear growth of the gate count with $m \cdot n$. However, unlike the qubit count, the total number of gates increases with $\ell$. This is attributed to the unary encoding and Dicke state preparation steps, which require a number of gates that scales linearly with $\ell$~\cite{patamawisut2025quantumcircuitdesigndecoded}. Additionally, the number of gates increases linearly with the number of decoder iterations $T$, which is expected since each iteration introduces a fixed number of additional gates to the circuit.

Both the total number of qubits and the total number of gates required by the DQI circuit exceed the capabilities of currently available quantum hardware. However, we observe that the quantum resource requirements scale sublinearly with the size of the max-XORSAT matrix $B$, quantified as $m \cdot n$. This suggests that, once sufficiently large quantum computers become available, the quantum circuit implementation proposed in this work could offer an efficient approach to solving industrially relevant ILP problems.

\subsection{Small instance circuit simulations}

\begin{figure}[h]
    \centering
    \includegraphics[width=0.7\linewidth]{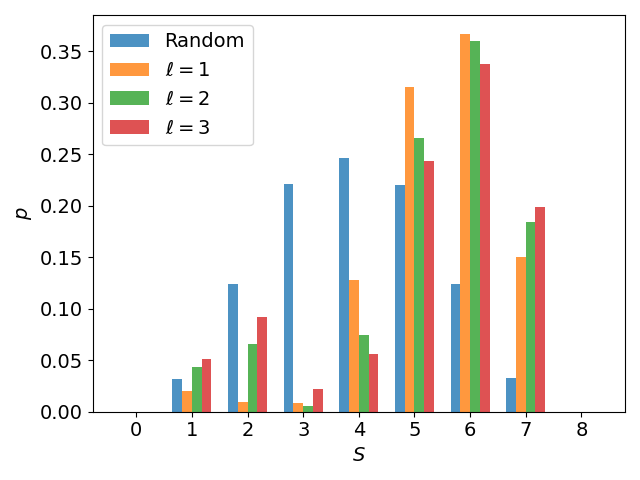}
    \caption{Histogram of the number of satisfied constraints obtained from DQI sampling for different values of $\ell$ and $T=1$, compared with uniform random sampling. Each sampling uses $10^4$ shots; note that for DQI, the number of usable bitstrings may be slightly lower due to postprocessing with decoding.}
    \label{fig: histogram DQI}
\end{figure}

In this section, we present the results of simulating the quantum circuit execution of the DQI algorithm for small problem instances. Figure~\ref{fig: histogram DQI} shows the complete histogram of the number of satisfied constraints obtained from sampling the output of the DQI circuit for different values of $\ell$. We compare these results to those obtained by sampling uniformly random bit-strings. The instance tested corresponds to the max-XORSAT problem defined by:

\begin{equation}
\begin{aligned}
B &= 
\begin{bmatrix}
1 & 1 & 0 & 0 & 0 & 0 \\
1 & 0 & 0 & 0 & 1 & 0 \\
0 & 1 & 1 & 0 & 0 & 0 \\
0 & 0 & 1 & 1 & 0 & 0 \\
0 & 0 & 0 & 1 & 0 & 1 \\
0 & 0 & 0 & 0 & 1 & 1 \\
0 & 0 & 1 & 0 & 0 & 1 \\
1 & 0 & 0 & 0 & 0 & 1
\end{bmatrix}, 
\quad
\mathbf{v} = 
\begin{bmatrix}
0 \\
0 \\
1 \\
1 \\
0 \\
0 \\
1 \\
1
\end{bmatrix}
\end{aligned}.
\end{equation}
This system leads to a DQI circuit consisting of 26 qubits when using $T=1$.

One can observe that the probability of obtaining the optimal solution (i.e., satisfying 7 constraints) is significantly higher when using DQI compared to random sampling. Furthermore, increasing the value of $\ell$ tends to improve the probability of finding the optimal solution—provided that the decoder's success rate is non-zero. Importantly, if we shift our focus from the probability of achieving the optimal solution to the average number of satisfied constraints, we find that the average is consistently below 7 across the different values of $\ell$ tested. This highlights an important subtlety: using only the expected number of satisfied constraints as a performance metric can underestimate the true performance of DQI. In particular, the expectation averages over all outcomes, including suboptimal ones, and therefore it underestimates the high likelihood of obtaining the optimal solution. 

Next, to cross‐verify both the DQI circuit implementation and our classical estimator of $\langle S\rangle$ that we use for larger instances, we generate three collections of small max-XORSAT problems and compare the empirical average number of satisfied constraints (with its sampling error) against the expected value computed via the method described in Sec.~\ref{sec: performance metrics}.  We generate random sparse max-XORSAT instances and run the DQI algorithm following the varied configurations specified in Section~\ref{sec: experiment design}. We test different configurations of the number of iterations and the maximum error $\ell$ to build confidence in both the quantum circuit implementation and the classical estimation method.

\begin{figure}[H]
  \centering
  \includegraphics[width=0.75\linewidth]{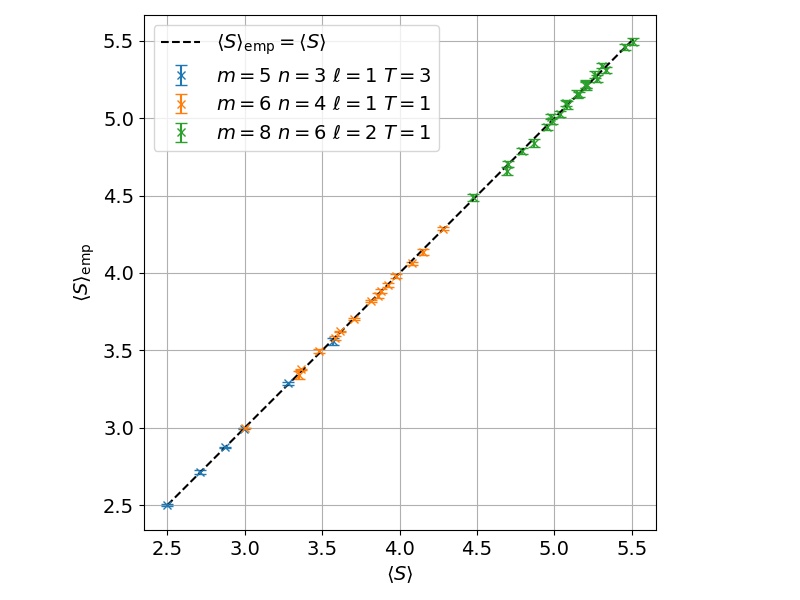}
  \caption{Comparison between predicted and empirical average number of satisfied constraints across three test regimes. Each point represents a different randomly sampled max-XORSAT instance. The plot shows excellent agreement between the analytical expected value and the empirical one we calculate by simulating our quantum circuit.}
  \label{fig: emp_vs_exp}
\end{figure}

In Fig.~\ref{fig: emp_vs_exp}, we compare the predicted and empirical values of $\langle S\rangle$ across all tested instances. Each point corresponds to a max-XORSAT problem, with vertical error bars indicating the sampling uncertainty in the empirical average. The dashed diagonal line represents perfect agreement between prediction and observation. We observe strong overall consistency between the theoretical estimates and the values obtained from quantum circuit sampling. To improve clarity and avoid excessive overlap, we do not plot all 50 random instances for each configuration. Instead, we display only the first occurrence of each unique predicted value of $\langle S\rangle$ within a given configuration. This prevents the accumulation of overlapping points that would otherwise obscure the visualization.

\section{Conclusion and Outlook}\label{sec:conclusions}

In this work, we provide a complete implementation of the DQI algorithm that can be applied to real-world industrial optimization problems once quantum hardware capabilities are available at scale. We show how any ILP can be transformed into a max-XORSAT instance, thereby extending the applicability of DQI beyond max-XORSAT problems. Our implementation includes a novel quantum circuit that coherently performs binary, hard-decision Belief Propagation. This quantum subroutine is used as the decoding step within the overall DQI quantum circuit we present.

We apply DQI to an industrially relevant use case: the optimization of configuration packages for vehicles. To this end, we formulate the problem as an ILP and provide a transformation into a max-XORSAT instance, which is then solved using DQI. Our results show that real-world optimization problems entail significant resource overhead in order to be made amenable to DQI, currently limiting its applicability to ``native" max-XORSAT problems or highly structured problems. Interestingly, we observe that all problem instances we constructed correspond to classical error-correcting codes with low distance ($d = 3$), which inherently constrains the optimal performance of the decoder, even under ideal conditions. Nonetheless, we find empirically that configuring the decoder to correct more errors than the code can reliably handle often leads to improved results.

At the problem sizes we are currently able to test, classical methods significantly outperform DQI in terms of performance. However, we observe favorable scaling of quantum resources for DQI, while classical methods are expected to exhibit exponential scaling with problem size. This suggests that, for sufficiently large instances, a crossover point between the performance of classical solvers like Gurobi and DQI could be possible. In our view, this keeps the question of quantum advantage for industrially relevant problems open.

Notably, our approach remains far from the asymptotic regime ($l/m \rightarrow \infty$) studied in~\cite{jordan2024optimizationdecodedquantuminterferometry}, which prevents us from leveraging the semicircle law described therein to predict the behavior of DQI for large-scale instances. This constrains our ability to make definitive claims about its performance in the large problem-size limit.

Additionally, our benchmarking of the two belief propagation algorithms used in this work shows that decoding performance degrades as the number of errors increases. This imposes a fundamental limitation on the average performance of DQI, even under ideal conditions. Furthermore, it is important to highlight a caveat of our benchmarking approach: we compare classical solvers, which return explicit solutions with a known number of satisfied clauses, to the expected number of satisfied constraints by DQI. Since the expected value of satisfied constraints is lower than the best individual sample, this comparison inherently underestimates the true performance of DQI. 

In future work, we aim to investigate several open questions that emerged from this study: 
\begin{enumerate}[label=\alph*.]
    \item Is it possible to engineer codes that maximize the code distance through an optimized transformation from ILP to max-XORSAT?
    \item Can the BP2 decoding algorithm be efficiently implemented as a quantum circuit—for example, by leveraging the techniques proposed in Refs.~\cite{Piveteau_2022,Renes_2017}?
    \item Are there industrially relevant problems that are naturally formulated as max-XORSAT, or that can be efficiently mapped to the polynomial intersection problem studied in Ref.~\cite{jordan2024optimizationdecodedquantuminterferometry}?
\end{enumerate}

\section{Acknowledgments} We would like to thank Hamed Mohammadbagherpoor and Norbert Widmann for insightful discussions.

\section*{Code}
The code accompanying this work is available on \url{https://bcg-x-official.github.io/dqi/}. 

\bibliography{bibliography}
\bibliographystyle{unsrt}

\section{Appendix}

\subsection{Benchmarking a Gauss-Jordan decoder}
\label{app: GJapp}
In this section, we test an implementation of a Gauss-Jordan decoder, similar to the one proposed in~\cite{patamawisut2025quantumcircuitdesigndecoded}, to evaluate its effectiveness in decoding errors from both square and rectangular matrices. The decoder under consideration is described in Algorithm~\ref{alg:GJ_decoder}. We verify that the decoder is always successful at correcting any number of errors when the parity-check matrix is a full-rank square matrix. In such cases, solving the max-XORSAT problem reduces to solving a system of linear equations.

\begin{figure}[h]
    \centering
    \includegraphics[width=0.9\linewidth]{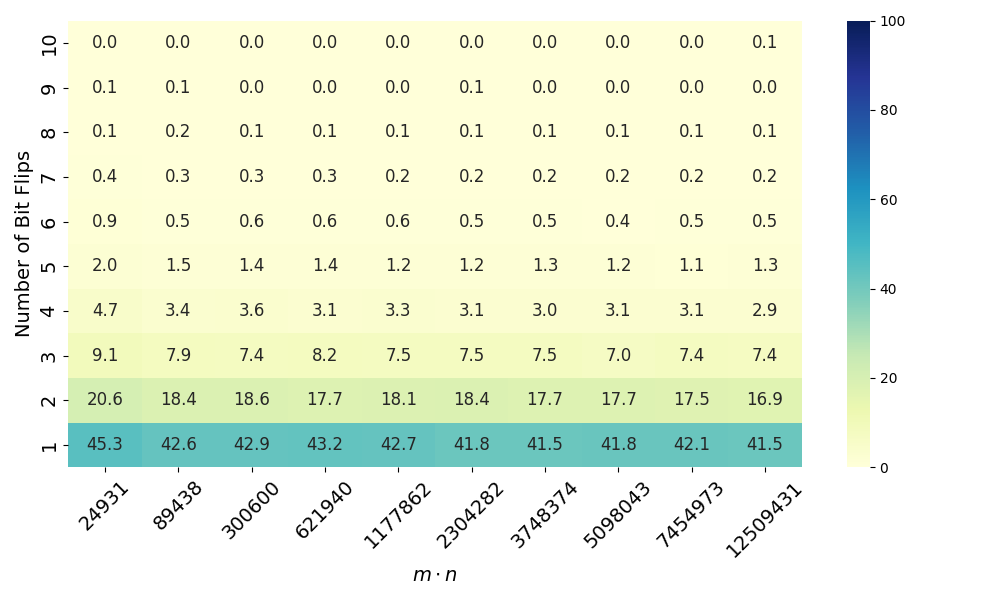}
    \caption{Success rate of the Gauss-Jordan decoder for the zero codeword for $\ell$ randomly generated  bit flips as a function of $\ell$ and the problem size, i.e, the size of $B$. We use $10^4$ sampled code words for each value of $\ell$ and $B$. }
    \label{fig:benchmark_GJ}
\end{figure}

Figure~\ref{fig:benchmark_GJ} presents an estimate of the success rate when decoding the zero codeword $\mathbf{c}$ corrupted by $\ell$ random bit flips. The plot shows how success varies with both the number of bit-flips and the problem size, which we define as the total number of entries in the parity-check matrix $B^T$. The results correspond to the same set of random ILP instances mapped to max-XORSAT problems that were previously used to benchmark the BP1 decoder in Fig.~\ref{fig:benchmark_gallager_bp}. We observe that for all tested problem sizes and numbers of bit flips, the BP1 decoder consistently outperforms the Gauss-Jordan decoder.
  
 \begin{algorithm}[h]
\caption{Gauss–Jordan Decoder }
\begin{algorithmic}[1]
\STATE \textbf{Input:} Parity check matrix $H \in \{0,1\}^{m \times n}$, received codeword $y \in \{0,1\}^n$
\STATE \textbf{Output:} Decoding success flag \texttt{success}, decoded bits vector \texttt{decoded\_bits}

\STATE $\texttt{syndrome} \gets (H \cdot y) \bmod 2$
\STATE $\texttt{H\_aug} \gets$ augmented matrix $[H~|~\texttt{syndrome}]$
\STATE $m \gets$ number of rows of $H$
\STATE $n \gets$ number of columns of $H$
\STATE $\texttt{row} \gets 0$
\STATE Initialize empty list \texttt{pivot\_cols}

\FOR{$\texttt{col} = 0$ to $n-1$}
    \STATE $\texttt{ones} \gets \{i ~|~ \texttt{H\_aug}[i, \texttt{col}] = 1 \text{ for } i \geq \texttt{row}\}$
    \IF{$\texttt{ones}$ is empty}
        \STATE \textbf{continue}
    \ENDIF
    \STATE $\texttt{pivot} \gets$ first index in $\texttt{ones} + \texttt{row}$
    \IF{$\texttt{pivot} \neq \texttt{row}$}
        \STATE Swap rows $\texttt{H\_aug}[\texttt{row}] \leftrightarrow \texttt{H\_aug}[\texttt{pivot}]$
    \ENDIF
    \STATE Append $\texttt{col}$ to \texttt{pivot\_cols}
    \FOR{each $i = 0$ to $m-1$ with $i \neq \texttt{row}$}
        \IF{$\texttt{H\_aug}[i, \texttt{col}] = 1$}
            \STATE $\texttt{H\_aug}[i] \gets \texttt{H\_aug}[i] \oplus \texttt{H\_aug}[\texttt{row}]$
        \ENDIF
    \ENDFOR
    \STATE $\texttt{row} \gets \texttt{row} + 1$
    \IF{$\texttt{row} = m$}
        \STATE \textbf{break}
    \ENDIF
\ENDFOR

\STATE Initialize $\texttt{e} \gets$ zero vector of length $n$
\FOR{$i = 0$ to $\texttt{len}(\texttt{pivot\_cols}) - 1$}
    \STATE $\texttt{e}[\texttt{pivot\_cols}[i]] \gets \texttt{H\_aug}[i, -1]$
\ENDFOR

\STATE $\texttt{decoded\_bits} \gets (y + e) \bmod 2$
\IF{all entries of \texttt{decoded\_bits} are zero}
    \STATE \textbf{return} (\texttt{True}, \texttt{decoded\_bits})
\ELSE
    \STATE \textbf{return} (\texttt{False}, \texttt{decoded\_bits})
\ENDIF
\end{algorithmic}
\label{alg:GJ_decoder}
\end{algorithm}

\subsection{Random parity check matrix generation}\label{app:sampling_B}
Due to the complications of testing the DQI algorithm for problems coming from industrial applications, i.e., even the smallest industrial problem is too big for detailed calculation, we have opted for sampling smaller instances of the parity check matrix $B^T$ in a way that it resembles the structure of a ILP transformed into a max-XORSAT instance. 

Following \cite{jordan2024optimizationdecodedquantuminterferometry}, we have that the degree of a constraint $i$ is given by $\sum_jB_{ij}$, i.e., the number of variables which appear in that constraint, with distribution $\kappa_{i'}$, i.e., the fraction of constraints having degree $i'$. Conversely, we have that the degree of a variable $j$ is given by $\sum_iB_{ij}$ with distribution $\Delta_{j'}$, meaning the fraction of variables with degree $j'$. In order to sample a random matrix $B_s\in\{0,1\}^{m\times n}$ from given distributions $\kappa$ and $\Delta$, we first sample arrays $c\sim\kappa$ and $r\sim\Delta$ so that $\sum_i c_i=\sum_j r_j$. This ensures that the total number of nonzero entries in $B_s$ is compatible for both marginal sums. The sampling is done via reject sampling and becomes costly for large matrix sizes. Then we use the greedy procedure described in Algorithm \ref{alg:initializing_random_B} to initialize a matrix $B_s$ so that the marginal sums are given by $c$ and $r$. Finally, we use a simple implementation of the \emph{swap algorithm}, which implements a Markov chain over the $B$ matrix, \cite{10.1007/978-1-4612-0801-3_3} described in Algorithm \ref{alg:swap}, which samples random instances of matrix $B$ fulfilling the marginal sum requirements.

\begin{algorithm}
\caption{Greedy Initial Matrix Generation from Row and Column Sums}
\begin{algorithmic}[1]
\REQUIRE Row sum vector $\texttt{row\_sums}$ of length $m$, column sum vector $\texttt{col\_sums}$ of length $n$
\ENSURE Binary matrix $B \in \{0,1\}^{m \times n}$ satisfying the degree constraints

\STATE $m \gets \texttt{length}(\texttt{row\_sums}),\quad n \gets \texttt{length}(\texttt{col\_sums})$
\STATE Initialize $B \gets$ zero matrix of size $m \times n$
\STATE $\texttt{rows} \gets$ indices of $\texttt{row\_sums}$ sorted in descending order of values
\STATE $\texttt{cols} \gets$ indices of $\texttt{col\_sums}$ sorted in descending order of values

\FOR{each $(i, r)$ in $\texttt{rows}$}
    \STATE $\texttt{possible} \gets$ current $\texttt{cols}$ sorted by descending sum
    \FOR{each $(j, c)$ in $\texttt{possible}$}
        \IF{$r > 0$ and $c > 0$ and $B[i, j] = 0$}
            \STATE $B[i, j] \gets 1$
            \STATE $\texttt{row\_sums}[i] \gets \texttt{row\_sums}[i] - 1$
            \STATE $\texttt{col\_sums}[j] \gets \texttt{col\_sums}[j] - 1$
            \STATE Update $\texttt{cols} \gets$ list of $(j, \texttt{col\_sums}[j])$ for all $j$
            \STATE $r \gets r - 1$
            \IF{$\texttt{row\_sums}[i] = 0$}
                \STATE \textbf{break}
            \ENDIF
        \ENDIF
    \ENDFOR
\ENDFOR

\STATE \textbf{return} $B$
\end{algorithmic}\label{alg:initializing_random_B}
\end{algorithm}

\begin{algorithm}
\caption{Swap Algorithm}
\begin{algorithmic}[1]

\STATE \textbf{Input:} Initial binary matrix $B_0 \in \{0,1\}^{m \times n}$, number of iterations $T$
\STATE \textbf{Output:} Modified binary matrix $B_s$

\FOR{$t = 1$ to $T$}
    \STATE Uniformly sample two distinct row indices $i_1, i_2$ and two distinct column indices $j_1, j_2$
    \STATE Let $S \gets$ the $2 \times 2$ submatrix of $B_{t-1}$ with rows $i_1, i_2$ and columns $j_1, j_2$
    \IF{$S = \begin{bmatrix}1 & 0\\ 0 & 1\end{bmatrix}$}
        \STATE Swap $S$ to:
        \STATE \hspace{1em} 
        $\begin{bmatrix}0 & 1\\ 1 & 0\end{bmatrix}$
        \STATE Set $B_t \gets B_{t-1}$ with the submatrix $S$ replaced by the swapped version
    \ELSE
        \STATE $B_t \gets B_{t-1}$ \COMMENT{No change}
    \ENDIF
\ENDFOR

\STATE \textbf{return} $B_s = B_T$
\end{algorithmic}\label{alg:swap}
\end{algorithm}

\end{document}